\documentclass[sn-mathphys-num]{sn-jnl}

\usepackage{bm}

\usepackage{booktabs}
\usepackage[normalem]{ulem}

\usepackage{hyperref}

\usepackage{acronym}
\acrodef{POD}[POD]{{Proper Orthogonal Decomposition}}
\acrodef{VAE}[VAE]{{Variational Autoencoder}}
\acrodef{AE}[AE]{{ Autoencoder}}
\acrodef{PCA}[PCA]{{Principal Component Analysis}}
\acrodef{ELBO}[ELBO]{{Evidence Lower Bound}}
\acrodef{KL}[KL]{ Kullback–Leibler}
\acrodef{ISOMAP}[ISOMAP]{Isometric Feature Mapping}
\acrodef{KNN}[KNN]{K-nearest Neighbors}
\acrodef{CNN}[CNN]{Convolutional Neural Network}
\acrodef{ROM}[ROM]{Reduced Order Model}

\usepackage{tikz}
\usepackage[dvipsnames]{xcolor}
\usepackage{overpic}

\definecolor{AOA20}{HTML}{F0C000}
\definecolor{AOA30}{HTML}{FF7700}
\definecolor{AOA40}{HTML}{FF0000}
\definecolor{AOA50}{HTML}{8B0000}
\definecolor{AOA60}{HTML}{1C1C1C}
\newcommand{\roundedline}[1]{%
  \tikz[baseline=-0.5ex] \draw[#1, line width=3pt, line cap=round] (0,0) -- (2.5mm,0);
}
\usepackage{graphicx}%
\usepackage{multirow}%
\usepackage{amsmath,amssymb,amsfonts}%
\usepackage{amsthm}%
\usepackage{mathrsfs}%
\usepackage[title]{appendix}%
\usepackage{xcolor}%
\usepackage{textcomp}%
\usepackage{manyfoot}%
\usepackage{booktabs}%
\usepackage{algorithm}%
\usepackage{algorithmicx}%
\usepackage{algpseudocode}%
\usepackage{listings}%
\usepackage{scalerel}
\usepackage{tikz}
\usetikzlibrary{svg.path}
\usepackage{cleveref}
\definecolor{orcidlogocol}{HTML}{A6CE39}
\tikzset{
  orcidlogo/.pic={
    \fill[orcidlogocol] svg{M256,128c0,70.7-57.3,128-128,128C57.3,256,0,198.7,0,128C0,57.3,57.3,0,128,0C198.7,0,256,57.3,256,128z};
    \fill[white] svg{M86.3,186.2H70.9V79.1h15.4v48.4V186.2z}
                 svg{M108.9,79.1h41.6c39.6,0,57,28.3,57,53.6c0,27.5-21.5,53.6-56.8,53.6h-41.8V79.1z M124.3,172.4h24.5c34.9,0,42.9-26.5,42.9-39.7c0-21.5-13.7-39.7-43.7-39.7h-23.7V172.4z}
                 svg{M88.7,56.8c0,5.5-4.5,10.1-10.1,10.1c-5.6,0-10.1-4.6-10.1-10.1c0-5.6,4.5-10.1,10.1-10.1C84.2,46.7,88.7,51.3,88.7,56.8z};
  }
}

\newcommand\orcidicon[1]{\href{https://orcid.org/#1}{\mbox{\scalerel*{
\begin{tikzpicture}[yscale=-1,transform shape]
\pic{orcidlogo};
\end{tikzpicture}
}{|}}}}

\usepackage{hyperref}

\theoremstyle{thmstyleone}%
\theoremstyle{thmstyletwo}%

\theoremstyle{thmstylethree}%

\begin{document}

\title[Article Title]{Information decomposition for disentangled and interpretable manifold learning of fluid flows via variational autoencoders}


\author*[1]{\fnm{Zhiyuan} \sur{Wang}}\email{zwang@ing.uc3m.es
\orcidicon{0000-0002-0020-5722}}

\author[1]{\fnm{Iacopo} \sur{Tirelli}}\email{iacopo.tirelli@uc3m.es \orcidicon{0000-0001-7623-1161}}
\author[1]{\fnm{Stefano} \sur{Discetti}}\email{sdiscett@ing.uc3m.es \orcidicon{0000-0001-9025-1505}}

\author[1]{\fnm{Andrea} \sur{Ianiro}}\email{aianiro@ing.uc3m.es \orcidicon{0000-0001-7342-4814}}

\affil*[1]{\orgdiv{Department of Aerospace Engineering}, \orgname{Universidad Carlos III de Madrid}, \orgaddress{\street{Avda. Universidad 30}, \city{Leganés}, \postcode{28911}, \state{Madrid}, \country{Spain}}}


\abstract{
We introduce an information-theoretic framework that uses variational autoencoders (VAEs) to extract compact, physically interpretable manifolds from high-dimensional flow-field data. To this end, the Kullback--Leibler (KL) divergence in the variational objective is decomposed into three complementary information-theoretic terms: the index-code mutual information, the total correlation, and the dimension-wise KL divergence. These terms explicitly regulate data compression, latent disentanglement, and geometric regularization. This establishes a principled basis for targeted latent-space design, allowing enhanced interpretability without sacrificing information capacity, a common drawback of heavily regularized VAE variants.
The approach is evaluated on two synthetic unsteady flow datasets. First, we consider a flow around a cylinder in a channel with variable cylinder position, diameter, and Reynolds number. Later, we also consider the flow around a NACA 0012 airfoil at varying angles of attack and subjected to strong vortex gusts with variable intensity, position, and length scale. Comparisons with Principal Component Analysis, Isometric Feature Mapping, and $\beta$-VAE demonstrate clear advantages in disentanglement and physical interpretability. The learned latent coordinates successfully separate distinct physical effects. Moreover, the proposed method demonstrates strong robustness to variations in the loss-weighting parameters, despite involving a larger number of such parameters.
}




\maketitle

\section{Introduction}
\label{sec:introduction}

In this study, we propose an information-theoretic approach for constructing compact and physically interpretable low-dimensional manifolds of unsteady flows using \acp{AE}.
Turbulent flows are inherently high-dimensional and nonlinear, making direct analysis, modeling, and control particularly challenging. 
Behind the apparent chaos, one can identify prominent, organized, and relatively low-dimensional patterns, often referred to as coherent structures\cite{Jiménez2018JFM}. Constructing low-dimensional coordinates that capture these coherent structures is an opportunity to build \acp{ROM}. An effective \ac{ROM} technique should yield physically consistent, parsimonious, and compact representations.
This need has motivated the development of manifold-learning and dimensionality-reduction techniques in fluid mechanics, which provide the basis for efficient  modeling, sensing and control of complex fluid flows\cite{Rowley2017annualreview}.

Linear statistical methods are widely used across various fields of applied science to identify and rank system coordinates based on the variance they represent in the data. In this context, a classical and widely used technique is \ac{PCA}. Originally developed in multivariate statistics \cite{pearson1901principal,hotelling1936simplified}, it appears under different names across disciplines: the Karhunen–Loève transform in mathematics \cite{schmidt1907theorie}, Empirical Orthogonal Functions in meteorology and climatology\cite{kutzbach1967empirical}, and \ac{POD} in fluid mechanics \cite{lumley1967atmospheric,Lumley1993annualreview}. \ac{PCA} reduces large datasets by expressing them in terms of principal components (modes) that are ranked according to the amount of variance they capture. Owing to its conceptual simplicity and broad applicability, \ac{PCA} has become the cornerstone of many data-driven analyses. Numerous extensions and modifications have also been developed to expand its capabilities and address its limitations~\cite{scholkopf1997kernel,jolliffe2016principal,mendez2023linear,tirelli2025meshless}.
Since this technique aims to compress the data while preserving as
much information as possible, it falls in the category of linear \ac{AE} \cite{baldi1989neural}. An autoencoder consists of an encoder–decoder architecture that maps high-dimensional input data to a low-dimensional latent representation and reconstructs the input from this representation.
Due to its linear nature, \ac{PCA} can struggle to capture strongly nonlinear flow features that do not lie in a single linear subspace. Consequently, often, a single coherent structure requires several modes to be represented. For instance, a travelling wave requires two modes with a phase shift of $90^\circ$, seldom including additional modes for higher-order harmonics (see, e.g., \citet{raiola2016wake}); however, the whole traveling wave should be described by one single fundamental coordinate, i.e., the phase angle of the traveling wave.
This limitation has motivated the development of nonlinear manifold-learning methods. 

Manifold learning methods, unlike \acp{AE}, seek low-dimensional representations that preserve as much as possible some metric of similarity, rather than compress the data. The underlying hypothesis here, the so-called \textit{manifold hypothesis}, is that data tend to lie near a low-dimensional hypersurface, the manifold.
A well-established manifold-learning algorithm is \ac{ISOMAP}\cite{tenenbaum2000global}, which extends classical multidimensional scaling by preserving geodesic distances on the manifold. In fluid mechanics, \citet{Farzamnik2023jfm} combined an \ac{ISOMAP} encoding of the snapshots data on a low-dimensional manifold with a decoding based on a locally-linear interpolation approach among the \ac{KNN}. By accounting for the intrinsic curved geometry of attractors or solution manifolds, \ac{ISOMAP} has demonstrated an enhanced ability to represent flows with strong nonlinear features when compared to \ac{PCA}. \citet{Luigi2024JFM} demonstrated that non-linear manifold learning can even describe complex controlled flows under different exogenous inputs. 
Despite its effectiveness, it is worth noting that \ac{ISOMAP} is built upon the computation of a geodesic distance matrix, which leads to some limitations. Firstly, the criterion is purely geometric and does not incorporate any physical constraints or invariances. As a result, the low-dimensional embedding may not capture the true system dynamics, and the reduced variables may not correspond to meaningful physical quantities. 
Moreover, \ac{ISOMAP} does not provide an explicit out-of-sample mapping: new states are not naturally embedded or decoded and typically require additional approximations (e.g., \ac{KNN}-based local interpolation as in \citet{Farzamnik2023jfm}) rather than a genuine learned mapping. In the framework of \citet{Farzamnik2023jfm}, a reliable reconstruction of the neighbor graph requires sufficiently dense sampling of the state space, which is not always available. Combined with the poor scalability due to the curse of dimensionality in pairwise distance computation, this may limit its applicability in high-dimensional turbulent flows.

In recent years, the rise of deep neural networks has introduced powerful new tools for nonlinear dimensionality reduction in fluid mechanics\cite{Brunton2020annualreview,zhang2025scientometric}. Among them, deep \acp{AE} stand out for their strong expressive power and reconstruction accuracy. 
The pioneering work by \citet{baldi1989neural} first demonstrated the potential of such architectures for dimensionality reduction.
The work by \citet{milano2002neural} is one of the first attempts to encode and decode a turbulent flow with a neural network, demonstrating improved reconstruction capabilities with respect to \acp{PCA}. 
Exploiting the field nature of fluid-flow snapshots \citet{lee2020model} demonstrated the potential of convolutional autoencoders.
Classical deep \acp{AE} learn an encoding-decoding mapping that is optimal for data compression and reconstruction but typically leave the latent space largely unconstrained, which may yield entangled and poorly interpretable coordinates. This issue can be alleviated by leveraging the flexibility of deep \acp{AE} in terms of model architecture and loss-function design, which has opened the way to a wide spectrum of variants tailored to different modeling objectives and application domains \cite{mienye2025deep}.

\citet{fukami2023grasping} recently proposed an observable-augmented convolutional autoencoder for manifold learning of fluid flows, in which a branch network maps the latent variables to auxiliary physical quantities, such as the lift coefficient; the corresponding prediction error is added to the loss function. In this way, the latent space is more strongly regularized and organized according to physically meaningful observables. Since its introduction, this approach has been successfully applied to a variety of flow problems\cite{FukamiTaira2025JFM, Fukami2025jfm}. Compared with generic unsupervised dimensionality reduction, this approach can be considered as a semi-supervised information–extraction process, balancing the prediction of target quantities with the reconstruction of the original fields.

While keeping an unsupervised training, probabilistic latent-variable framework can be adopted to impose distributional and information-theoretic constraints on the latent space. In the machine learning community, the seminal work of \citet{kingma2014autoencoding} marked a paradigm shift by introducing a probabilistic framework for generative modeling, enabling the synthesis of new data instances. This approach led to the development of the \acp{VAE}. Unlike standard \acp{AE}, which learn a deterministic mapping to an unconstrained latent space, \acp{VAE} place an explicit probabilistic model, typically Gaussian, on the latent variables, making them naturally suited for uncertainty quantification and generation of new data with a decoder. 
Training a \ac{VAE} consists of maximizing the \ac{ELBO}, implemented as a loss function that balances data reconstruction accuracy with a \ac{KL} divergence enforcing consistency between the learned latent distribution and a prescribed prior. The introduction of an adjustable hyperparameter $\beta$ to promote disentanglement in the latent space gave rise to one of the most widely used \ac{VAE} variants, the $\beta$-\ac{VAE} \cite{higgins2017beta}. By weighting the \ac{KL} divergence term, $\beta$ provides a mechanism to balance reconstruction accuracy against latent-space disentanglement: values of $\beta > 1$ encourage disentanglement, whereas $\beta < 1$ prioritizes reconstruction fidelity. \citet{solera2024beta} demonstrated how effectively this class of \acp{VAE} can be applied for reduced-order modeling in fluid flows. Recent works have also demonstrated the applicability of this framework in challenging aerodynamic regimes, such as transonic flows~\citep{kang2022physics}, as well as for complex geometries like supercritical airfoils~\citep{wang2021flow,li2022physically}.

Despite these advances, significant challenges remain in applying manifold-learning and autoencoder-based methods to fluid flows. For observable-augmented autoencoders, the semi-supervised learning process relies on the availability of well-defined target quantities. In complex flows, however, identifying suitable observables and dealing with their practical measurability and measurement uncertainty both require substantial, case-dependent effort. \acp{VAE}, by contrast, provide a probabilistic framework in which the latent space is explicitly modeled as a distribution, making it natural to encode and quantify uncertainty\cite{mousavi2025low}. Nevertheless, the \ac{KL} term in the \ac{ELBO} couples multiple effects, and simply increasing its weight to ``improve regularization'' effectively forces the latent variables toward an independent standard normal distribution, leading to information capacity loss, latent-space posterior collapse, and associated training instabilities\cite{Lucas2019DontBlameELBO}.

To address these issues, we propose an \ac{ELBO}-decomposed \ac{VAE} framework for manifold learning of fluid flows. The idea of decomposing the \ac{ELBO} arises from the deep generative modeling literature\cite{kim2018disentangling, chen2018isolating}. In this work, we build on the \ac{ELBO}–TC decomposition proposed by \citet{chen2018isolating} and explicitly split the \ac{KL} term into three components with clear information-theoretic interpretations. This decomposition enables more targeted control of latent-space properties and mitigates the trade-offs and information loss induced by the coupled terms in the original $\beta$-\ac{VAE}. It is worth remarking that the ELBO decomposition does not impose a manifold hypothesis. However, the ELBO-decomposed VAE enables the compactness and disentanglement of low-dimensional coordinates, which are thus consistent with the underlying physical manifold of the flow data.

The remainder of this paper is organized as follows. Section~\ref{sec:method} introduces the proposed methodology, including the network architecture and training details. It also describes the datasets used for validation. These include (i) the wake of a circular cylinder in a channel, with variations in cylinder position, diameter, and Reynolds number, and (ii) the flow around a NACA~0012 airfoil at different angles of attack, subjected to strong vortex gusts with varying intensity, position, and length scale. Section~\ref{sec:results} compares the proposed method with classical and state-of-the-art manifold-learning and dimensionality-reduction approaches on these datasets, focusing on the learned latent-space structure and reconstruction accuracy. Finally, Section~\ref{sec:conclusions} concludes the paper and discusses implications and future directions.

\section{Methodology}
\label{sec:method}
\subsection{ELBO-TC decomposition for variational autoencoders}
\label{sec:elbo-decomp}

\begin{figure}
\includegraphics[width=\textwidth]{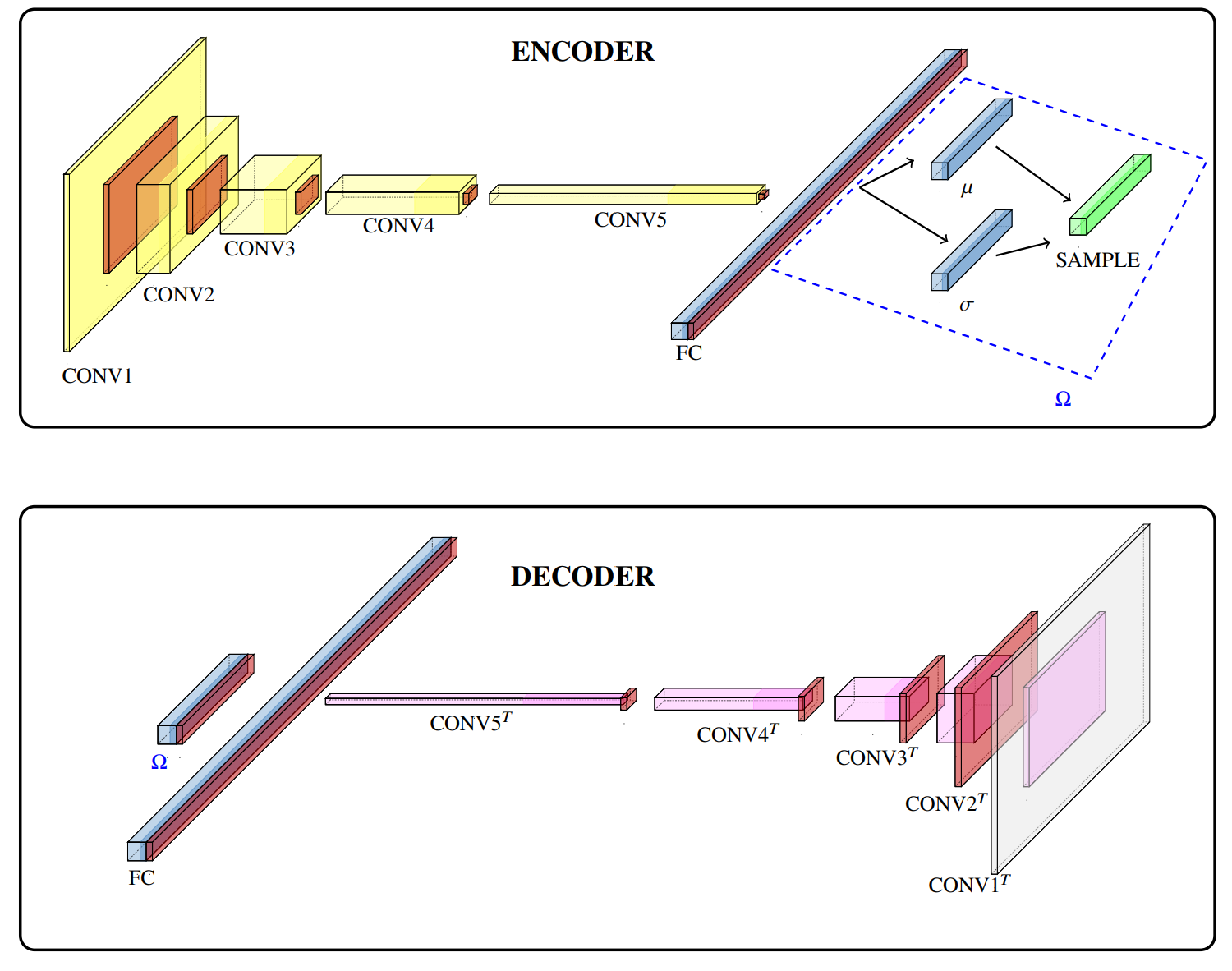}

\caption{Schematic of the network architecture. Top: encoder with five convolutional layers (CONV\#, stride 2, in yellow), activation function (GELU, in red), fully connected layer (FC, in blue). Two parallel layers ($\mu$, $\sigma$) estimate the mean and variance of the latent distribution, which is sampled during training (in green) to produce the decoder input ($\Omega$). Bottom: decoder with $\Omega$, an FC layer, and five transposed convolutional layers (CONV\#$^T$, in pink).}
\label{Fig:schemtaic}
\end{figure}

Following the ELBO-TC decomposition introduced in~\citet{chen2018isolating}, we adopt this information-theoretic objective and apply it to manifold learning for fluid flows via \acp{VAE}. This approach explicitly isolates, and thereby independently controls, different contributions to the loss function, allowing them to be tailored to the specific objectives of the application.

Consider a dataset of flow snapshots $\{\bm{x}_n\}_{n=1}^{N}$, where $\bm{x}_n\in\mathbb{R}^{d_x}$ denotes the $n$-th snapshot and $d_x$ is the number of spatial dimensions. \ac{AE} can be viewed as an encoding-decoding mapping with an encoder $f_{\phi}$ mapping high-dimensional discretized flow fields to a low-dimensional set of latent coordinates $\bm{z} = f_{\phi}(\bm{x})$, and a decoder reconstructing the full field from this compact representation $\hat{\bm{x}} = g_{\theta}(\bm{z})$. The symbols $\phi$ and $\theta$ 
denote the trainable parameters of the encoder and decoder, respectively, e.g. weights and biases of the underlying neural 
networks. 

As stated in the introduction, a \ac{VAE} extends this framework by adopting a probabilistic latent description. Specifically, the latent variable $\bm{z}\in\mathbb{R}^{d_z}$ is endowed with a prior distribution $p(\bm{z})$, and the decoder defines a conditional generative model $p_\theta(\bm{x}\mid\bm{z})$ parameterized by $\theta$, enabling both reconstruction and generation of physically plausible flow fields from latent samples. Since the true posterior $p_\theta(\bm{z}\mid\bm{x}_n)$ is intractable, it is approximated by a variational encoder distribution $q_\phi(\bm{z}\mid\bm{x}_n)$ with parameters $\phi$, which maps each snapshot to a distribution in latent space. The standard \ac{VAE} is trained by minimizing the negative evidence lower bound (\ac{ELBO}):

\begin{equation}
\begin{aligned}
\mathcal{L}_{\mathrm{VAE}}
&= \frac{1}{N} \sum_{n=1}^{N}
\Big(
-\mathbb{E}_{q_\phi(\bm{z }\mid \bm{x }_n)}[\log p_\theta(\bm{x}_n \mid \bm{z})]
\\
&\qquad\qquad
+ \mathrm{KL}\big(q_\phi(\bm{z} \mid \bm{x}_n)\,\|\,p(\bm{z})\big)
\Big).
\end{aligned}
\label{eq:vae-elbo}
\end{equation}

 In practice, $p_\theta(\bm{x}\mid\bm{z})$ is modeled as an isotropic Gaussian; under this assumption, the term $-\mathbb{E}_{q_\phi(\bm{z}\mid\bm{x}_n)}[\log p_\theta(\bm{x}_n\mid\bm{z})]$ reduces (up to an additive constant and a scaling factor) to an $L_2$ reconstruction error.
$\mathrm{KL}(\cdot\|\cdot)$ denotes the \ac{KL} divergence between the variational posterior $q_\phi(\bm{z}\mid\bm{x}_n)$ and the prior $p(\bm{z})$.
The first term in Eq.~\eqref{eq:vae-elbo} promotes reconstruction accuracy of
the data, while the second term regularizes the latent representation
towards the prior.

\citet{chen2018isolating} proposed a further decomposition of the \ac{KL} term.
Let $n \sim p(n)$ denote a uniformly sampled data index (i.e., sampling snapshots with equal weight in the loss function), with the empirical index distribution $p(n)=1/N$. This corresponds to weighting all snapshots equally in the loss function. For notational convenience, we write $q_\phi(\bm{z}\mid n) \equiv q_\phi(\bm{z}\mid \bm{x}_n)$.
With the factorized prior $p(\bm{z})=\prod_{j=1}^{d_z} p(\bm{z}_j)$, the index-averaged \ac{KL} divergence can be rewritten as:
\begin{equation}
\begin{aligned}
\mathbb{E}_{p(n)}\!\Big[
  \mathrm{KL}\big(q_\phi(\bm{z} \mid n)\,\|\,p(\bm{z})\big)
\Big]
&=
\frac{1}{N}\sum_{n=1}^N
\mathrm{KL}\big(q_\phi(\bm{z}\mid n)\,\|\,p(\bm{z})\big)
\\
&=
\mathrm{KL}\big(q_\phi(\bm{z},n)\,\|\,q_\phi(\bm{z})\,p(n)\big)
\\
&\quad+
\mathrm{KL}\Big(q_\phi(\bm{z})\,\Big\|\,\prod_{j} q_\phi(z_j)\Big)
\\
&\quad+
\sum_{j} \mathrm{KL}\big(q_\phi(z_j)\,\|\,p(z_j)\big),
\end{aligned}
\label{eq:elbo-tc}
\end{equation}

Here $q_\phi(\bm{z},n)$ is the joint distribution over the latent variable $\bm{z}$ and the dataset index $n$. 
The marginal $q_\phi(\bm{z})=\sum_{n=1}^N p(n)\,q_\phi(\bm{z}\mid n)$ is the aggregated posterior.
Finally, $q_\phi(\bm{z}_j)=\int q_\phi(\bm{z})\,\mathrm{d}\bm{z}_{\setminus j}$ denotes the marginal distribution of the $j$-th latent component, where $\bm{z}_{\setminus j}$ collects all latent dimensions except $\bm{z}_j$.

The three terms on the right-hand side of Eq.~\eqref{eq:elbo-tc} have
clear information-theoretic interpretations:
\begin{itemize}
  \item $\mathrm{KL}\big(q_\phi(\bm{z},n)\,\|\,q_\phi(\bm{z})\,p(n)\big)$ is the
        \emph{index--code mutual information} (MI). It represents the
        mutual information between  the data
        variable and latent variable based on the empirical data distribution $q(\bm{z},n)$. This term quantifies how much snapshot-specific information (e.g., instantaneous small-scale fluctuations, intermittent events, or phase details) is retained in the latent coordinates. Slightly penalizing this mutual information enforces an information bottleneck, encouraging a more compact representation that preferentially captures the dominant flow structures. In practice, its weight is often kept at unity, consistent with the standard \ac{VAE} objective. 
        
\item $\mathrm{KL}\Big(q_\phi(\bm{z})\,\Big\|\,\prod_{j} q_\phi(\bm{z}_j)\Big)$ is the
      \emph{total correlation} (TC) of the latent variables. It measures
      the statistical dependence among latent dimensions (the \emph{disentanglement}). Reducing TC encourages the aggregated posterior $q(\bm{z})$ to factorize across dimensions, i.e., to make latent factors approximately independent. For fluid flows, this favors latent coordinates that isolate distinct physical degrees of freedom, thereby improving interpretability. 

\item $\sum_{j} \mathrm{KL}\big(q_\phi(\bm{z}_j)\,\|\,p(\bm{z}_j)\big)$ is the
      \emph{dimension-wise KL divergence} (Dim-KL). It penalizes deviations of each
      marginal latent distribution $q(\bm{z}_j)$ from its prior $p(\bm{z}_j)$, thereby
      enforcing prior matching on a per-dimension basis. The role of this term is twofold: first, it regularizes the latent distribution to
      prevent it from overcomplexity or distortion; second, it keeps each latent dimension well-aligned with the prior, improving sampling-based generation. However, for flow-field data with highly complex, non-Gaussian structure, overly strong matching to a simple prior can induce severe information-capacity loss; in coupled objectives such as the $\beta$-\ac{VAE} loss, this trade-off is largely unavoidable. In contrast, the \ac{ELBO}-TC decomposition allows this issue to be mitigated by relaxing the Dim-KL term independently.

\end{itemize}

Note that the decomposition of the KL divergence involves the aggregated posterior and related marginals, whose exact computation is intractable. To solve this computational issue, as proposed by \citet{chen2018isolating}, one can adopt minibatch stratified sampling (MSS).

Once identified the components of Eq.~\eqref{eq:elbo-tc}, one can define a loss function  as in Eq.~\eqref{eq:vae-elbo-weight} that can tune the relative importance of mutual information between data and latent variables, disentanglement, and dimension-wise prior matching, thus shaping the information content and statistical structure of the latent space:

\begin{equation}
\begin{alignedat}{2}
\mathcal{L}_{\mathrm{DKL\text{-}VAE}} &= \frac{1}{N} \sum_{n=1}^{N} \big\{
- \mathbb{E}_{q_\phi(\mathbf{z} \mid \mathbf{x}_n)} \big[\log p_\theta(\mathbf{x}_n \mid \mathbf{z})\big] \\
&\quad + \lambda_{\mathrm{MI}} \, \mathrm{KL}\big(q_\phi(\mathbf{z},n) \,\|\, q_\phi(\mathbf{z})\,p(n)\big) \\
&\quad + \lambda_{\mathrm{TC}} \, \mathrm{KL}\Big(q_\phi(\mathbf{z}) \,\Big\|\, \prod_j q_\phi(z_j)\Big) \\
&\quad + \lambda_{\mathrm{Dim\text{-}KL}} \sum_j \mathrm{KL}\big(q_\phi(z_j) \,\|\, p(z_j)\big)
\big\},
\end{alignedat}
\label{eq:vae-elbo-weight}
\end{equation}

\noindent where the weights $\lambda_{\mathrm{MI}}$, $\lambda_{\mathrm{TC}}$ and $\lambda_{\mathrm{Dim-KL}}$ refer to the corresponding terms of the loss.

In the remainder of this paper, we refer to the proposed method as the Decomposed-KL VAE (DKL-VAE). 

\begin{table*}[t]
\centering
\caption{Summary of datasets used for validation.}
\label{tab:dataset-details}
\renewcommand{\arraystretch}{1.2}
\begin{tabular}{p{3.2cm} p{4.2cm} p{4.2cm}}
\hline
\textbf{} & \textbf{Cylinder} & \textbf{Airfoil} \\
\hline
Flow configuration
& $2D$ incompressible channel flow past a circular cylinder, Poiseuille velocity profile 
& $2D$ flow around a NACA$0012$ airfoil with incoming large-scale vortical gusts \\

Flow regimes
& Steady laminar wakes and unsteady vortex shedding
& Transient, extreme aerodynamic response \\

Geometric parameters
& Cylinder radius $R/H\in[0.05,\,0.2]$ m; center $(x_c,y_c)$ with
$x_c/H\in[0.375,\,1.25]$ m, $y_c/H\in[0.25,\,0.75]$ m; channel width $H=0.4$ 
& Airfoil chord $c$; angle of attack $\alpha\in[20^\circ,30^\circ,40^\circ,50^\circ,60^\circ]$ \\

Reynolds number
& $\mathrm{Re}=u_{\mathrm{in}}^{\max}(2R)/\nu \in [10,\,200]$
& $\mathrm{Re}=U_{\infty}c/\nu = 100$ \\

Additional parameters
& Inflow velocity $u_{\mathrm{in}}^{\max}\in[0.5,\,2.25]$ m/s
& Gust: strength $G\in[-4,4]$, length scale $L\in[0.5,2]$,
 vertical offset $y_0/c\in[-0.5,0.5]$ \\

Input
& Two-component velocity field $[u,v]$, $32\times 64\times 2$ 
& Spanwise vorticity field $\omega_z$, $64\times 128\times 1$ \\

Number of cases           & $1036$            & $155$ \\
    Snapshots per case        & $50$              & $238$            \\
    Total snapshots (frames)  & $51800$           & $36890$          \\

\hline
\end{tabular}
\end{table*}

In practical applications, one typically needs to (i) determine an
appropriate latent dimensionality and (ii) identify which latent coordinates are most informative once a model is trained. Both can be addressed naturally within the VAE framework. Redundant latent dimensions tend to collapse during optimization, either by becoming nearly deterministic (vanishing posterior variance) or collapsing to the prior, yielding an uninformative latent space that carries no data-dependent information and thus contributes near-zero to the \ac{KL}. This suggests a simple strategy: starting from a conservatively large latent dimension and then inferring the effective dimensionality by inspecting the variance and KL. Then the model is retrained with the latent dimension adjusted to this estimated effective dimensionality.

To this end, we can also rank the latent dimensions using a KL-based activity score derived from the original
regularizer $\mathrm{KL}\!\big(q_\phi(\bm{z}\mid x_n)\,\|\,p(\bm{z})\big)$, as analyzed and validated in the literature\cite{KaaeNIPS2016}. Specifically, we quantify the \emph{activity} of the $j$-th latent
component $a_j$ by the expected one-dimensional KL divergence:
\begin{equation}
  a_j
  = \mathbb{E}_{p(n)}\!\Big[
      \mathrm{KL}\big(q_\phi(\bm{z}_j \mid n)\,\|\,p(\bm{z}_j)\big)
    \Big],
  \label{eq:latent-activity}
\end{equation}
which admits a closed form under the assumption of standard Gaussian prior and 
diagonal Gaussian encoder. In particular, assuming $p(\bm{z}_j)=\mathcal{N}(0,1)$
and $q_\phi(\bm{z}_j\mid \bm{x}_n)=\mathcal{N}\!\big(\mu_j(\bm{x}_n),\sigma_j^2(\bm{x}_n)\big)$:

\begin{equation}
  \mathrm{KL}\big(q_\phi(z_j\mid \bm{x}_n)\,\|\,p(z_j)\big)
  = \frac{1}{2}\Big(\mu_j(\bm{x}_n)^2+\sigma_j^2(\bm{x}_n)-\log\sigma_j^2(\bm{x}_n)-1\Big).
  \label{eq:kl-1d-closed}
\end{equation}

To quantitatively evaluate the statistical dependence between latent variables and physical parameters, together with the dependence across different latent dimensions, we employ the Hilbert--Schmidt Independence Criterion (HSIC)~\cite{gretton2005measuring}. HSIC is a kernel-based dependence measure that can capture both linear and nonlinear associations between two variables. Given two random variables $X$ and $Y$, the empirical HSIC is defined as:
\begin{equation}
\mathrm{HSIC}(X,Y) = \frac{1}{(n-1)^2} \mathrm{tr}(\mathbf{K}\mathbf{H}\mathbf{L}\mathbf{H}),
\end{equation}
where $n$ is the number of samples, $\mathbf{K}$ and $\mathbf{L}$ are the kernel (Gram) matrices constructed from the samples of $X$ and $Y$, respectively, and $\mathbf{H}=\mathbf{I}-\frac{1}{n}\mathbf{1}\mathbf{1}^\top$ denotes the centering matrix. In this work, to facilitate comparison across different variable pairs, we use the normalized HSIC, written as:
\begin{equation}
\mathrm{nHSIC}(X,Y) =
\frac{\mathrm{HSIC}(X,Y)}
{\sqrt{\mathrm{HSIC}(X,X)\,\mathrm{HSIC}(Y,Y)}}.
\end{equation}
With this normalization, the dependence strength can be compared more consistently across different latent dimensions and physical parameters. 
A larger normalized HSIC value indicates a stronger statistical dependence, while a smaller value suggests weaker dependence or greater independence.

\subsection{Details of the architecture}
The architecture used in this paper is summarized in Fig.~\ref{Fig:schemtaic}. The overall design is consistent across test cases, while the input resolution and channel widths are adapted to each dataset. The encoder is a convolutional stack with stride $2$, which progressively reduces the spatial resolution and extracts multiscale flow features. The resulting feature maps are flattened and passed through a fully connected layer to form a compact representation. Two parallel output heads, each of size $d_z$, parameterize the latent distribution by predicting its mean and variance. During training, latent samples are drawn via the reparameterization trick and provided to the decoder. For the cylinder dataset, the encoder comprises three convolutional layers with channel widths $(16,32,64)$ and takes inputs of size $(32,64)$; the fully connected layer has width $128$. For the airfoil dataset, the encoder comprises four convolutional layers with channel widths $(12,24,48,96)$ and takes inputs of size $(64,128)$; the fully connected layer has width $256$.

The decoder mirrors the encoder in a nearly symmetric manner. Latent samples are first mapped through a fully connected layer and reshaped to match the encoder's final convolutional output shape. A sequence of transposed-convolution layers then progressively restores the spatial resolution while decreasing the number of channels. A final transposed-convolution layer with two output channels produces the reconstructed fields. The Gaussian Error Linear Unit (GELU) \cite{hendrycks2016gaussian} is used throughout, except for the output layer, where a linear activation is applied.

The network is optimized using Adam for up to $600$ epochs. A plateau-based learning-rate decay strategy is further adopted: when the validation loss fails to improve for $20$ consecutive epochs, the learning rate is reduced by a factor of $0.5$.

\begin{figure}[t]
  \centering
  \begin{overpic}[width=0.60\textwidth, unit = 1mm]{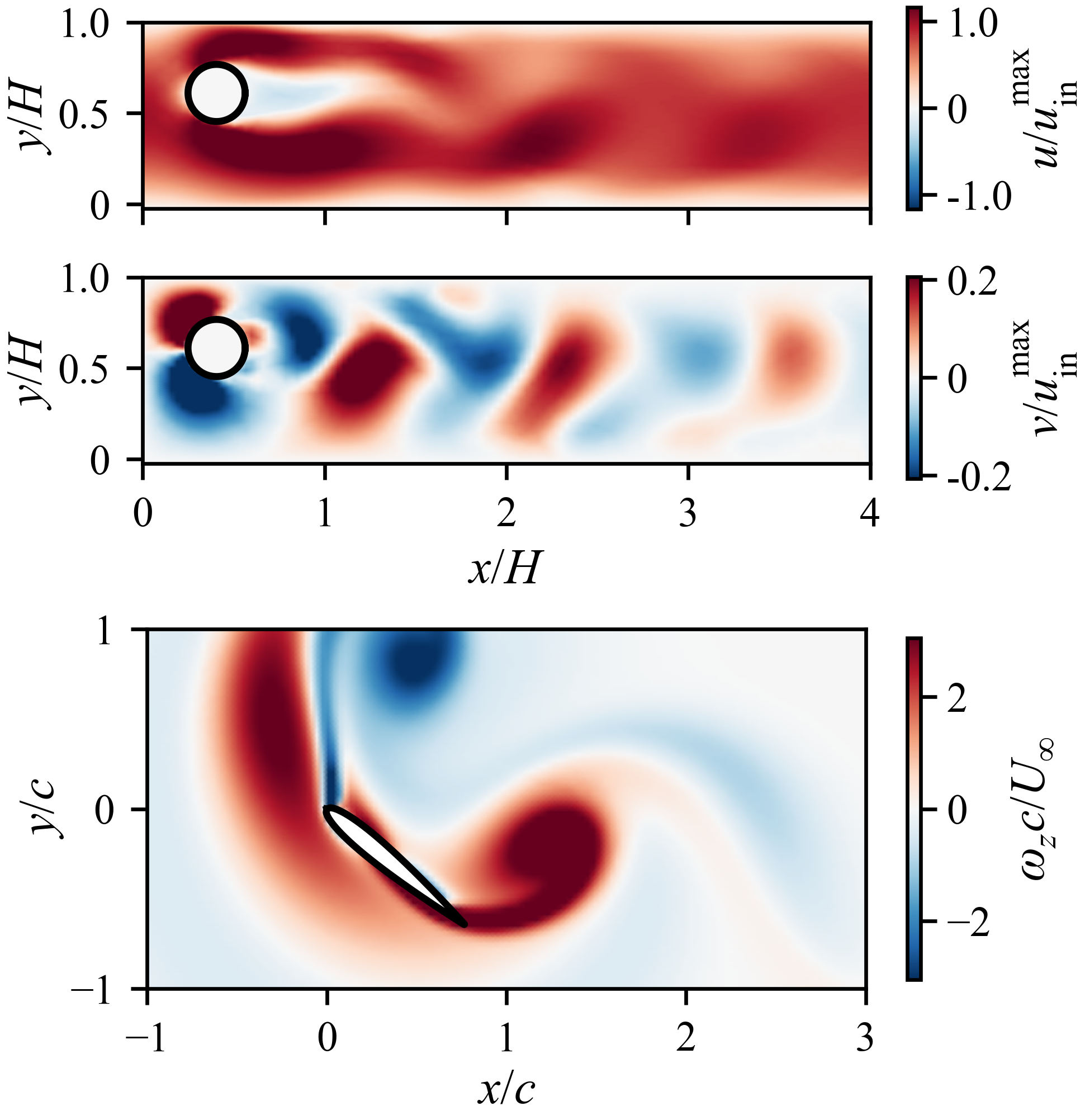}
  \put(-4,99){\parbox{10mm}{\centering {{a)}}}}
  \put(-4,76){\parbox{10mm}{\centering {{b)}}}}
  \put(-4,44){\parbox{10mm}{\centering {{c)}}}}
  \end{overpic}
  
  \caption{Representative flow snapshots from the 
  datasets. Top, cylinder testcase: a) Streamwise velocity component $u$, b) transverse velocity component $v$. Bottom, spanwise vorticity $\omega_z$ of the airfoil testcase. }
  \label{fig:datasets}
\end{figure}

\begin{figure*}[t]
  \centering
  \begin{overpic}[width=0.98\textwidth, unit = 1mm]{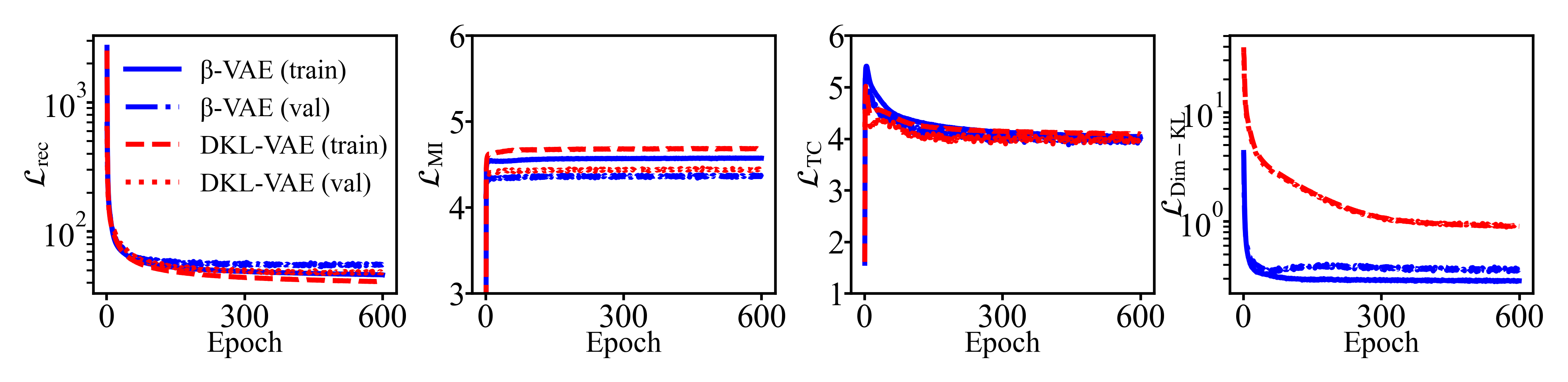}
  \put(0,23){\parbox{10mm}{\centering {{a)}}}}
  \put(23,23){\parbox{10mm}{\centering {{b)}}}}
  \put(47,23){\parbox{10mm}{\centering {{c)}}}}
  \put(72,23){\parbox{10mm}{\centering {{d)}}}}
  \end{overpic}
  
  \caption{Convergence histories of all loss terms on the cylinder dataset for $\beta$-VAE (blue, train continuous and validation dashed) with $\beta=15$ and DKL-VAE (red, train long dotted and validation dotted) with $\lambda_{\mathrm{TC}}=15, \lambda_{\mathrm{MI}}=\lambda_{\mathrm{Dim-KL}}=1$ . From left to right: reconstruction loss $\mathcal{L}_{rec}$, index–code mutual information  term loss $\mathcal{L}_{MI}$, total correlation term loss $\mathcal{L}_{TC}$ and  dimension-wise KL divergence term loss $\mathcal{L}_{Dim-KL}$. }
  \label{fig:cyl-loss-curves}
\end{figure*}

\begin{figure*}[t]
\vspace{5mm}
  \centering
  \begin{overpic}[width=0.95\textwidth, unit = 1mm]{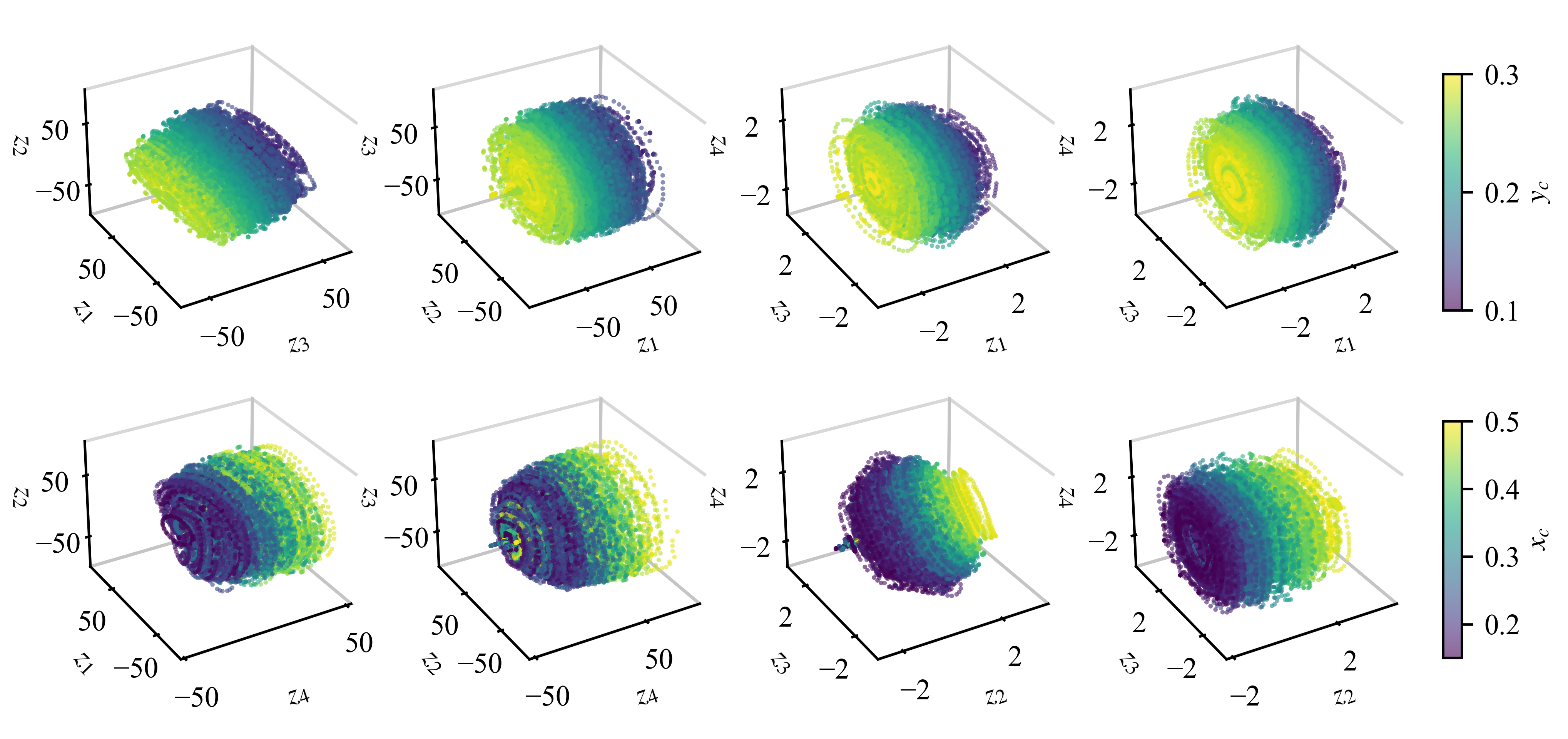}
  \put(-5,50){\parbox{45mm}{\centering \textbf{{PCA}}}}
  \put(29,50){\parbox{20mm}{\centering \textbf{{ISOMAP}}}}
  \put(51,50){\parbox{20mm}{\centering \textbf{{$\beta$-VAE}}}}
  \put(72,50){\parbox{20mm}{\centering \textbf{{DKL-VAE}}}}
  \put(1,44){\parbox{10mm}{ {{a.1)}}}}
  \put(1,23){\parbox{10mm}{ {{a.2)}}}}
  \put(24,44){\parbox{10mm}{ {{b.1)}}}}
  \put(24,23){\parbox{10mm}{ {{b.2)}}}}
  \put(46,23){\parbox{10mm}{{{c.2)}}}}
  \put(46,44){\parbox{10mm}{{{c.1)}}}}
  \put(68,23){\parbox{10mm}{ {{d.2)}}}}
  \put(68,44){\parbox{10mm}{ {{d.1)}}}}

  \end{overpic}
  
  \caption{Latent-space distributions on the cylinder dataset: (a) PCA, (b) ISOMAP, (c) $\beta$-VAE, and (d) DKL-VAE. Colored by:  spanwise cylinder center coordinate $y_c$ (top), streamwise cylinder center coordinate $x_c$ (bottom). For each method, latent dimensions are indexed according to the method-specific importance measure, and the embeddings are visualized in a common coordinate layout to facilitate an intuitive comparison.}
  \label{fig:cyl-latent-space}
\end{figure*}

\begin{figure*}[t]
  \centering
  \vspace{10mm} 
   \begin{overpic}[width=0.99\textwidth, unit = 1mm]{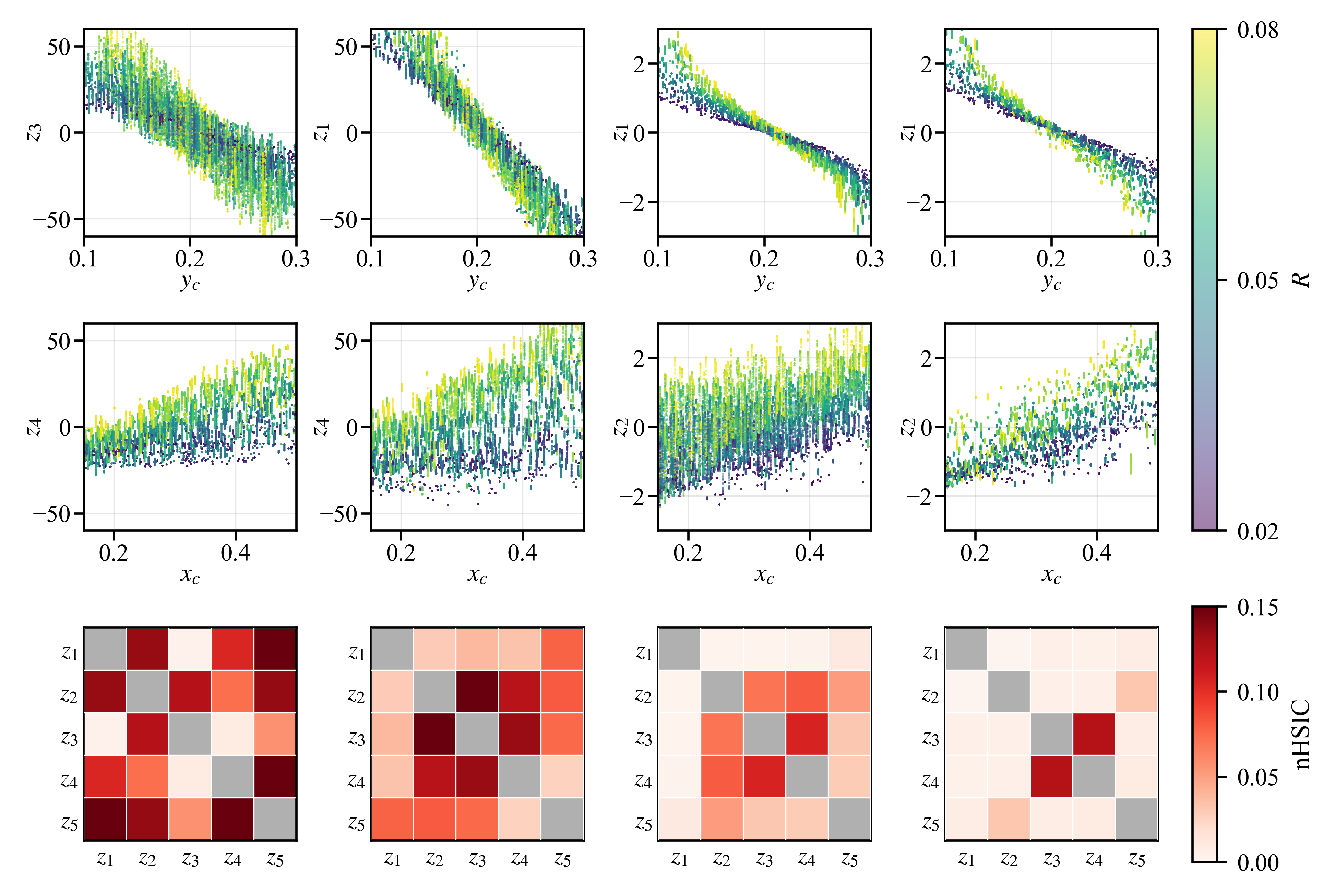}
  \put(11,68){\parbox{10mm}{\centering \textbf{{PCA}}}}
  \put(29.5,68){\parbox{20mm}{\centering \textbf{{ISOMAP}}}}
  \put(50,68){\parbox{20mm}{\centering \textbf{{$\beta$-VAE}}}}
  \put(72,68){\parbox{20mm}{\centering \textbf{{DKL-VAE}}}}
  \put(1,66){\parbox{10mm}{ {{a.1)}}}} 
  \put(23,66){\parbox{10mm}{ {{b.1)}}}}
  \put(45,66){\parbox{10mm}{{{c.1)}}}}
  \put(66,66){\parbox{10mm}{ {{d.1)}}}}
  
  \put(1,44){\parbox{10mm}{ {{a.2)}}}} 
  \put(23,44){\parbox{10mm}{ {{b.2)}}}}
  \put(45,44){\parbox{10mm}{{{c.2)}}}}
  \put(66,44){\parbox{10mm}{ {{d.2)}}}}
  
  \put(1,21){\parbox{10mm}{ {{a.3)}}}}
  \put(23,21){\parbox{10mm}{ {{b.3)}}}}
  \put(45,21){\parbox{10mm}{{{c.3)}}}}
  \put(66,21){\parbox{10mm}{ {{d.3)}}}}
\put(7,50){\parbox{25mm}{\tiny $\mathrm{nHSIC}=0.6480$}}
\put(28.5,50){\parbox{18mm}{\tiny $\mathrm{nHSIC}=0.8847$}}
\put(50,50){\parbox{18mm}{\tiny$\mathrm{nHSIC}=0.8503$}}
\put(71.5,50){\parbox{18mm}{\tiny $\mathrm{nHSIC}=0.8912$}}

\put(7,28){\parbox{18mm}{\tiny $\mathrm{nHSIC}=0.1581$}}
\put(28.5,28){\parbox{18mm}{\tiny $\mathrm{nHSIC}=0.0973$}}
\put(50,28){\parbox{18mm}{\tiny $\mathrm{nHSIC}=0.1062$}}
\put(71.5,28){\parbox{18mm}{\tiny $\mathrm{nHSIC}=0.3876$}}

  \end{overpic}
  
\caption{Latent-space analysis of the cylinder dataset using different dimensionality reduction methods: (a) PCA, (b) ISOMAP, (c) $\beta$-VAE, and (d) DKL-VAE, with nHSIC used to characterize the dependence between latent dimensions and physical parameters, as well as the dependencies among latent dimensions. Top: latent-space projections color-coded by the spanwise cylinder-center coordinate $y_c$; middle: latent-space projections color-coded by the streamwise cylinder-center coordinate $x_c$; bottom: nHSIC analysis among latent dimensions.}
  \label{fig:cyl-latent-alignment}
\end{figure*}

\begin{figure*}[t]
  \centering
   \vspace{10mm} 
   \begin{overpic}[width=1.1\textwidth, unit = 1mm]{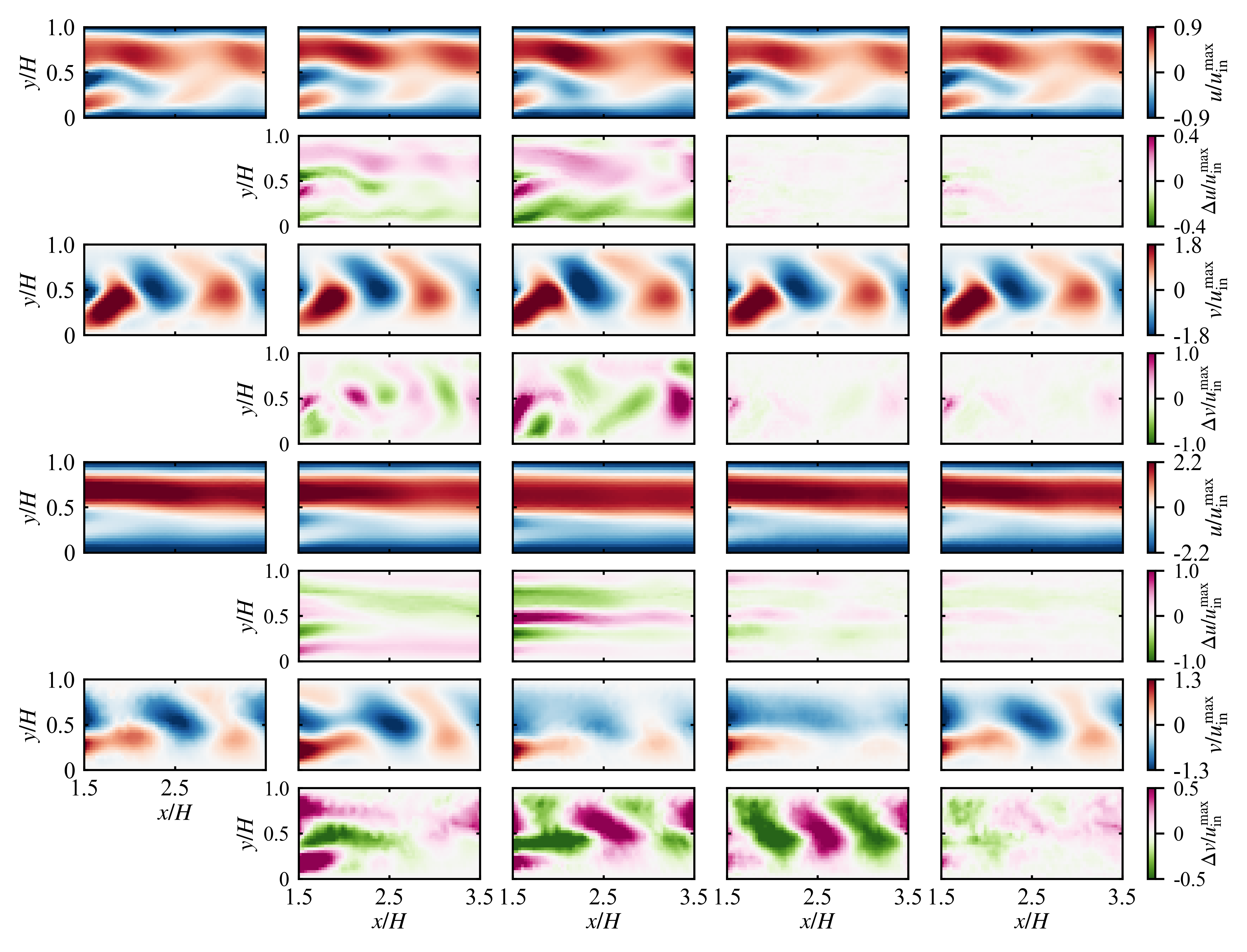}
  \put(9,76){\parbox{10mm}{\centering \textbf{{Reference}}}}
  \put(24,76){\parbox{20mm}{\centering \textbf{{PCA}}}}
    \put(41,76){\parbox{20mm}{\centering \textbf{{ISOMAP}}}}

  \put(58.5,76){\parbox{20mm}{\centering \textbf{{$\beta$-VAE}}}}
  \put(75.5,76){\parbox{20mm}{\centering \textbf{{DKL-VAE}}}}

\put(24.2,72.5){\parbox{10mm}{{a.1)}}}
\put(24.2,63.8){\parbox{10mm}{{a.2)}}}
\put(24.2,55.1){\parbox{10mm}{{a.3)}}}
\put(24.2,46.4){\parbox{10mm}{{a.4)}}}
\put(24.2,37.6){\parbox{10mm}{{a.5)}}}
\put(24.2,28.9){\parbox{10mm}{{a.6)}}}
\put(24.2,20.2){\parbox{10mm}{{a.7)}}}
\put(24.2,11.5){\parbox{10mm}{{a.8)}}}

\put(41.4,72.5){\parbox{10mm}{{b.1)}}}
\put(41.4,63.8){\parbox{10mm}{{b.2)}}}
\put(41.4,55.1){\parbox{10mm}{{b.3)}}}
\put(41.4,46.4){\parbox{10mm}{{b.4)}}}
\put(41.4,37.6){\parbox{10mm}{{b.5)}}}
\put(41.4,28.9){\parbox{10mm}{{b.6)}}}
\put(41.4,20.2){\parbox{10mm}{{b.7)}}}
\put(41.4,11.5){\parbox{10mm}{{b.8)}}}

\put(58.6,72.5){\parbox{10mm}{{c.1)}}}
\put(58.6,63.8){\parbox{10mm}{{c.2)}}}
\put(58.6,55.1){\parbox{10mm}{{c.3)}}}
\put(58.6,46.4){\parbox{10mm}{{c.4)}}}
\put(58.6,37.6){\parbox{10mm}{{c.5)}}}
\put(58.6,28.9){\parbox{10mm}{{c.6)}}}
\put(58.6,20.2){\parbox{10mm}{{c.7)}}}
\put(58.6,11.5){\parbox{10mm}{{c.8)}}}

\put(75.8,72.5){\parbox{10mm}{{d.1)}}}
\put(75.8,63.8){\parbox{10mm}{{d.2)}}}
\put(75.8,55.1){\parbox{10mm}{{d.3)}}}
\put(75.8,46.4){\parbox{10mm}{{d.4)}}}
\put(75.8,37.6){\parbox{10mm}{{d.5)}}}
\put(75.8,28.9){\parbox{10mm}{{d.6)}}}
\put(75.8,20.2){\parbox{10mm}{{d.7)}}}
\put(75.8,11.5){\parbox{10mm}{{d.8)}}}

  \end{overpic}
  \caption{Reconstructed snapshots of the cylinder dataset for different methods: (a) PCA, (b) ISOMAP, (c) $\beta$-VAE, and (d) DKL-VAE. For each test case, the reconstructed $u$- and $v$-velocity components are shown together with the corresponding error contours. Rows 1--4 correspond to the case $[u^{\max}_{\mathrm{in}}, x_c/H, y_c/H, R/H] = [2.200, 1.079, 0.392, 0.098]$, showing the reconstructed $u$ component, reconstructed $v$ component, and their corresponding error contours, respectively. Rows 5--8 correspond to the case $[u^{\max}_{\mathrm{in}}, x_c/H, y_c/H, R/H] = [1.243, 0.390, 0.421, 0.104]$, showing the reconstructed $u$ component, reconstructed $v$ component, and their corresponding error contours, respectively.}
  \label{fig:cyl-recon}
\end{figure*}

\begin{figure*}[t]
  \centering
  \begin{overpic}[width=0.98\textwidth, unit = 1mm]{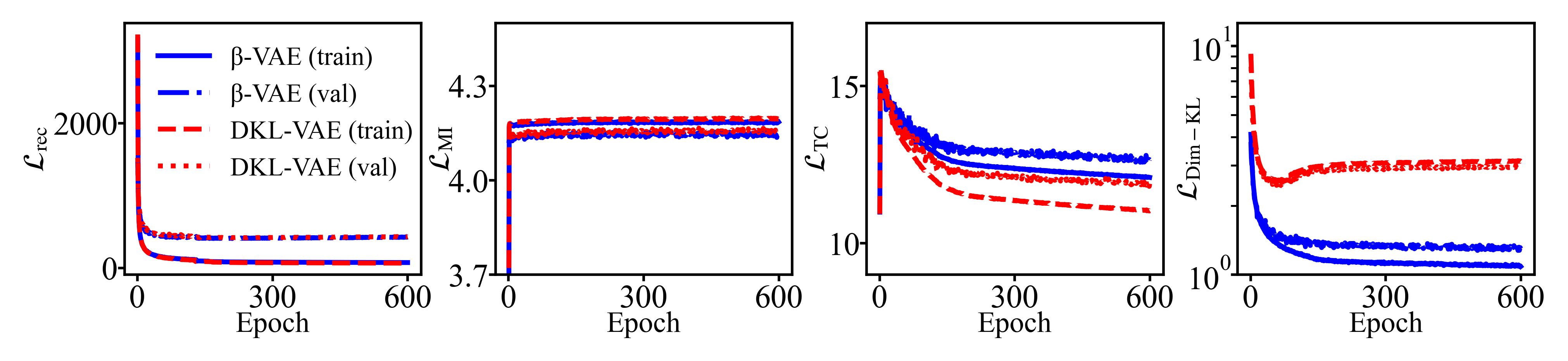}
  \put(2,21.5){\parbox{10mm}{\centering {{a)}}}}
  \put(25.5,21.5){\parbox{10mm}{\centering {{b)}}}}
  \put(49,21.5){\parbox{10mm}{\centering {{c)}}}}
  \put(72.5,21.5){\parbox{10mm}{\centering {{d)}}}}
  \end{overpic}
  
  \caption{Convergence histories of all loss terms on the airfoil dataset for $\beta$-VAE (blue, train continuous and validation dashed) with $\beta=10$ and DKL-VAE (red, train long dotted and validation dotted) with $\lambda_{\mathrm{TC}}=10$, $\lambda_{\mathrm{Dim-KL}}=4$ and $\lambda_{\mathrm{MI}}=1$. }
  \label{fig:Airfoil-loss-curves}
\end{figure*}

\begin{figure*}[t]
  \centering
  \begin{overpic}[width=0.95\textwidth, unit=1mm]{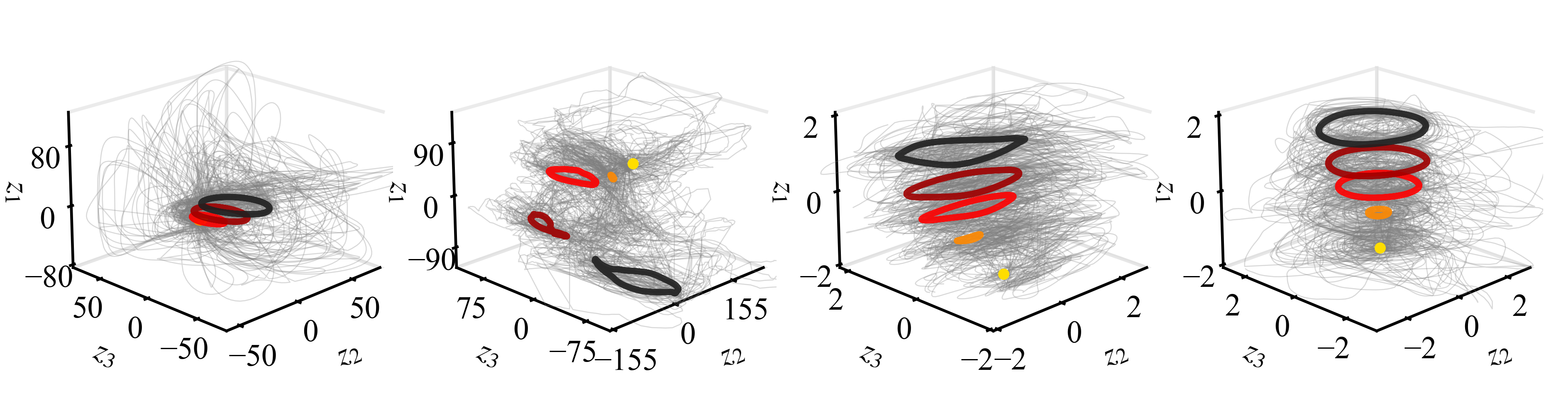}

    \put(11,24){\parbox{10mm}{\centering \textbf{PCA}}}
    \put(32,24){\parbox{20mm}{\centering \textbf{ISOMAP}}}
    \put(55,24){\parbox{20mm}{\centering \textbf{$\beta$-VAE}}}
    \put(81,24){\parbox{20mm}{\centering \textbf{DKL-VAE}}}

    \put(1,20.5){\parbox{10mm}{{a)}}}
    \put(26,20.5){\parbox{10mm}{{b)}}}
    \put(51,20.5){\parbox{10mm}{{c)}}}
    \put(76,20.5){\parbox{10mm}{{d)}}}

    \put(97,22){\parbox{20mm}{\centering {$\alpha[\text{deg}]$}}}

    \put(100,19){\roundedline{AOA20}\hspace{1mm}20}
    \put(100,16){\roundedline{AOA30}\hspace{1mm}30}
    \put(100,13){\roundedline{AOA40}\hspace{1mm}40}
    \put(100,10){\roundedline{AOA50}\hspace{1mm}50}
    \put(100,7){\roundedline{AOA60}\hspace{1mm}60}

  \end{overpic}

  \caption{Latent-space distributions on the airfoil dataset, from left to right: PCA, ISOMAP, $\beta$-VAE, and DKL-VAE. Undisturbed cases color-coded with the angle of attack; disturbed wake cases in light gray.}
  \label{fig:Airfoiol-latent-space}
\end{figure*}

\begin{figure*}[t]
  \centering
   \begin{overpic}[width=1\textwidth, unit = 1mm]{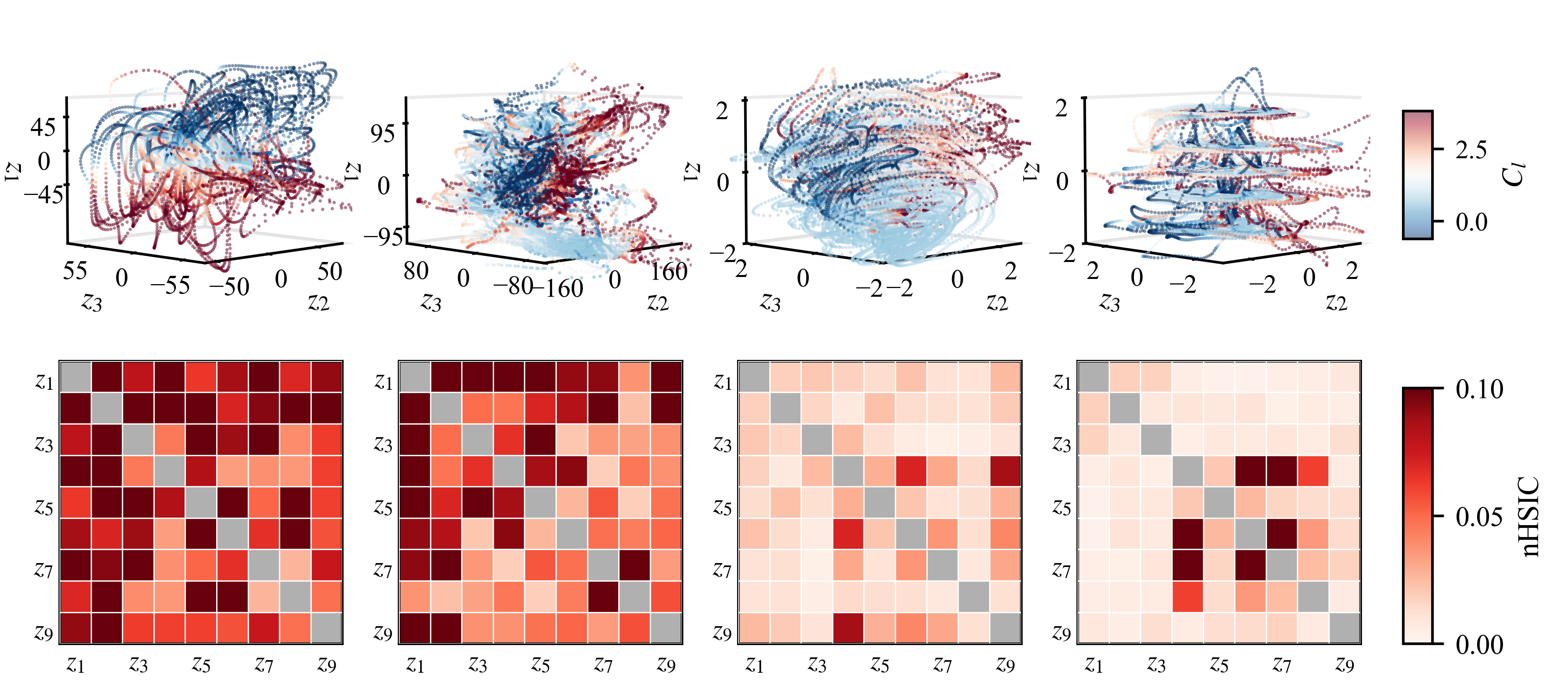}
  \put(10,43){\parbox{10mm}{\centering \textbf{{PCA}}}}
  \put(28,43){\parbox{20mm}{\centering \textbf{{ISOMAP}}}}
  \put(48,43){\parbox{20mm}{\centering \textbf{{$\beta$-VAE}}}}
  \put(70.5,43){\parbox{20mm}{\centering \textbf{{DKL-VAE}}}}

  \put(1,40){\parbox{10mm}{{a.1)}}}
  \put(22.5,40){\parbox{10mm}{{b.1)}}}
  \put(44.5,40){\parbox{10mm}{{c.1)}}}
  \put(66,40){\parbox{10mm}{{d.1)}}}

  \put(1,23){\parbox{10mm}{{a.2)}}}
  \put(22.5,23){\parbox{10mm}{{b.2)}}}
  \put(44.5,23){\parbox{10mm}{{c.2)}}}
  \put(66,23){\parbox{10mm}{{d.2)}}}

  \end{overpic}
  \caption{Latent-space analysis of the airfoil dataset for different dimensionality reduction methods: (a) PCA, (b) ISOMAP, (c) $\beta$-VAE, and (d) DKL-VAE. Top: latent-space projections color-coded by the lift coefficient $C_l$; bottom: nHSIC analysis showing the dependence among latent dimensions.}
  \label{fig:Airfoil-latent-alignment}
\end{figure*}

\begin{figure*}[t]
  \centering
  \vspace{5mm} 
   \begin{overpic}[width=1.1\textwidth, unit = 1mm]{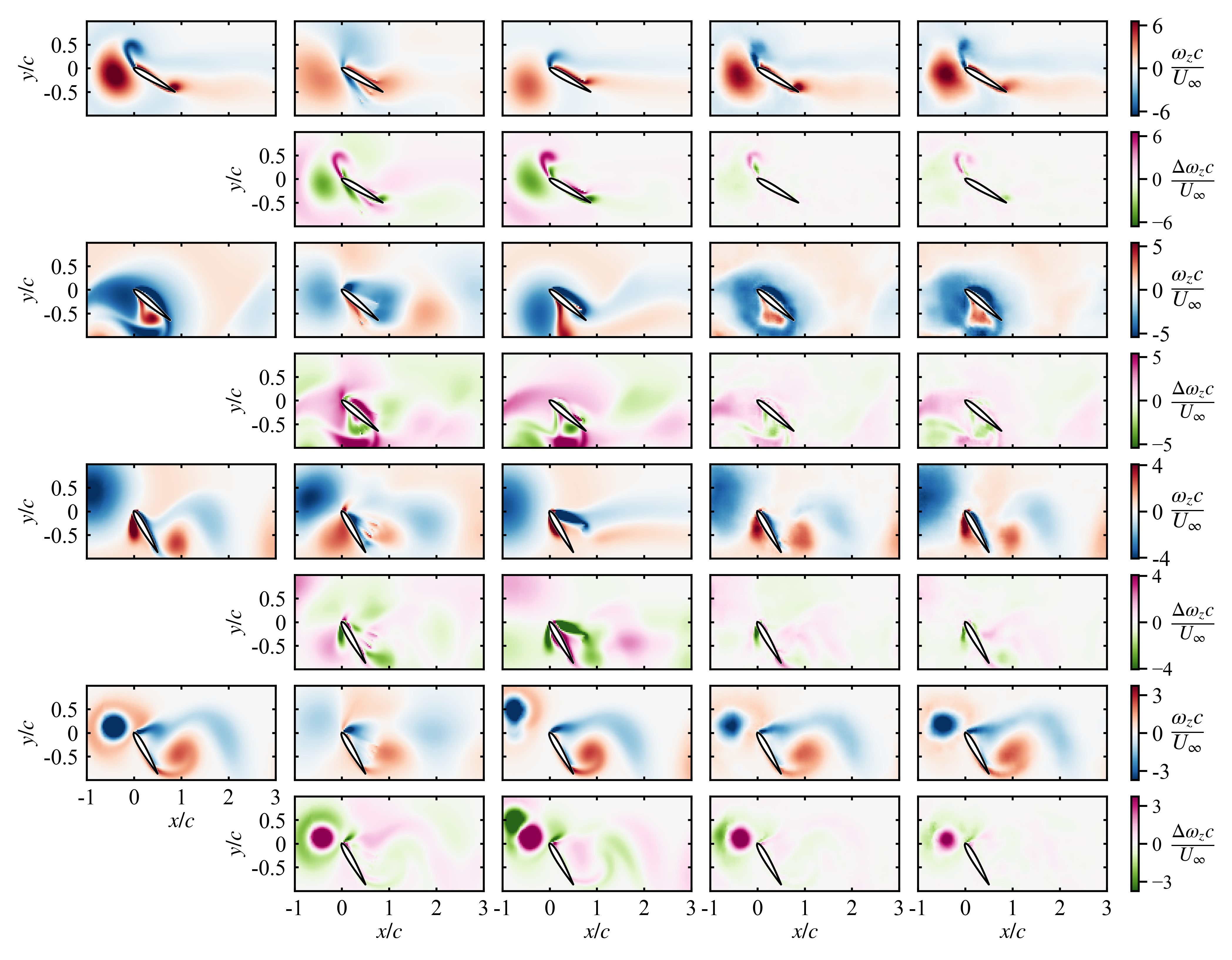}
  \put(9,78){\parbox{10mm}{\centering \textbf{{Reference}}}}
  \put(25,78){\parbox{20mm}{\centering \textbf{{PCA}}}}
    \put(42,78){\parbox{20mm}{\centering \textbf{{ISOMAP}}}}

  \put(59,78){\parbox{20mm}{\centering \textbf{{$\beta$-VAE}}}}
  \put(77,78){\parbox{20mm}{\centering \textbf{{DKL-VAE}}}}
\put(24.5,74.5){\parbox{10mm}{{a.1)}}}
\put(24.5,65.5){\parbox{10mm}{{a.2)}}}
\put(24.5,56.5){\parbox{10mm}{{a.3)}}}
\put(24.5,47.5){\parbox{10mm}{{a.4)}}}
\put(24.5,38.5){\parbox{10mm}{{a.5)}}}
\put(24.5,29.5){\parbox{10mm}{{a.6)}}}
\put(24.5,20.5){\parbox{10mm}{{a.7)}}}
\put(24.5,11.5){\parbox{10mm}{{a.8)}}}

\put(41.3,74.5){\parbox{10mm}{{b.1)}}}
\put(41.3,65.5){\parbox{10mm}{{b.2)}}}
\put(41.3,56.5){\parbox{10mm}{{b.3)}}}
\put(41.3,47.5){\parbox{10mm}{{b.4)}}}
\put(41.3,38.5){\parbox{10mm}{{b.5)}}}
\put(41.3,29.5){\parbox{10mm}{{b.6)}}}
\put(41.3,20.5){\parbox{10mm}{{b.7)}}}
\put(41.3,11.5){\parbox{10mm}{{b.8)}}}

\put(58.1,74.5){\parbox{10mm}{{c.1)}}}
\put(58.1,65.5){\parbox{10mm}{{c.2)}}}
\put(58.1,56.5){\parbox{10mm}{{c.3)}}}
\put(58.1,47.5){\parbox{10mm}{{c.4)}}}
\put(58.1,38.5){\parbox{10mm}{{c.5)}}}
\put(58.1,29.5){\parbox{10mm}{{c.6)}}}
\put(58.1,20.5){\parbox{10mm}{{c.7)}}}
\put(58.1,11.5){\parbox{10mm}{{c.8)}}}

\put(74.9,74.5){\parbox{10mm}{{d.1)}}}
\put(74.9,65.5){\parbox{10mm}{{d.2)}}}
\put(74.9,56.5){\parbox{10mm}{{d.3)}}}
\put(74.9,47.5){\parbox{10mm}{{d.4)}}}
\put(74.9,38.5){\parbox{10mm}{{d.5)}}}
\put(74.9,29.5){\parbox{10mm}{{d.6)}}}
\put(74.9,20.5){\parbox{10mm}{{d.7)}}}
\put(74.9,11.5){\parbox{10mm}{{d.8)}}}

  \end{overpic}
  \caption{
  Reconstructed snapshots of the airfoil dataset for different methods: (a) PCA, (b) ISOMAP, (c) $\beta$-VAE, and (d) DKL-VAE. For each test case, the reconstructed flow field is shown together with the corresponding error contour. Rows 1--2 correspond to the case $[G, L, y_0/c] = [3.4, 1.0, -0.3]$, showing the reconstructed snapshot and its error contour, respectively. Rows 3--4 correspond to the case $[G, L, y_0/c] = [-3.8, 2.0, -0.5]$, Rows 5--6 correspond to the case $[G, L, y_0/c] = [-3.0, 2.0, 0.5]$, and Rows 7--8 correspond to the case $[G, L, y_0/c] = [-3.6, 0.5, 0]$, again showing the reconstructed snapshot and its error contour, respectively.}
  
  \label{fig:Airfoil-recon}
\end{figure*}

\subsection{Datasets for validation}

The proposed methodology is validated on two benchmark flow datasets:
the wake of a cylinder-in-channel \cite{pfaff2021learning} and an airfoil
gust-encounter\cite{fukami2023grasping}.  Figure~\ref{fig:datasets} shows representative flow snapshots from both datasets. For the cylinder-in-channel dataset, the convolutional neural network takes the two in-plane velocity components, $u$ and $v$, as a two-channel input. For the airfoil
gust-encounter dataset, the input is the single-channel spanwise vorticity, $\omega_z$. All relevant specifications of the datasets used in our study are summarized in Tab.~\ref{tab:dataset-details}. 
The cylinder dataset consists of the $2D$ incompressible channel flows past a circular cylinder, where the inflow is prescribed by a Poiseuille-type (parabolic) velocity profile. The dataset is obtained from finite-element simulations performed in COMSOL. The dataset is parameterized by the maximum inlet velocity $u_{\mathrm{in}}^{\max}$ of each case, the cylinder radius $R$, and the cylinder center location $(x_c,y_c)$, inside a channel of width $H=0.4$m. Because the cylinder is placed in a finite-width channel, the flow is subject to the wall confinement effect. Moreover, the no-slip walls may induce secondary vorticity dynamics that interact with the cylinder wake, especially when the cylinder is close to the walls. Both effects are strongly dependent on the wall proximity, i.e., they vary with the radius $R$ and the center location $(x_c,y_c)$. Additional details about this dataset and the corresponding computational method are available in Ref.~\cite{pfaff2021learning}.

The gust-encounter dataset consists of $2D$ flow fields around a NACA$0012$ airfoil at $\mathrm{Re}=U_{\infty}c/\nu=100$, with $U_\infty$ being the freestream velocity and $c$ being the chord. The dataset was generated from direct numerical simulations using the immersed boundary method on $2.88 \times 10^{4}$ grid points. Gust–airfoil interactions are modeled by placing a large-scale vortex upstream of the
airfoil, which convects and impinges on the airfoil wake to produce transient, extreme aerodynamic responses. The dataset includes five angles of attack. Additional details about this dataset and the corresponding computational method are available in Ref.~\cite{fukami2023grasping}.

Although the datasets lie within a low-Reynolds-number regime, the coupling among multiple variables and physical effects induces sufficiently rich nonlinear dynamics, making them difficult to represent in a compact and physically-interpretable way. To capture 99\% of the total energy, classical PCA requires 29 and 179 modes for the two datasets, respectively, which indicates the complexity of the datasets For each $\alpha$, the gust is further parameterized by its strength $G = u_{\mathrm{gust}}/U_{\infty}$, length scale $L = D/c$, and vertical offset $y_0/c$, where $u_{\mathrm{gust}}$, $D$, and $y_0$ denote the maximum tangential velocity, diameter, and center position of the vortical gust, respectively. This yields 30 distinct operating conditions for each angle of attack.

In practical use, we apply several pre-processing steps before feeding the flow fields into the convolutional network. For the cylinder dataset, we interpolate the two-component velocity field $[u,v]$ onto a fixed downstream wake window, $x/H \in[1.5,\,3.5]$, to avoid the influence of the solid cylinder boundary on processing based on \acp{CNN}. Moreover, since the dataset includes simulations initialized from a uniform flow and thus contains a transient spin-up phase, we discard cases in which the flow does not reach a statistically steady regime, which typically occurs for very low inlet velocities or when the cylinder is placed too close to the walls. For the airfoil dataset, the raw data consist of vorticity fields computed by direct numerical simulation using an immersed-boundary method on a $120\times240$ grid. Because the immersed-boundary formulation yields values inside the solid body, we retain the full computational domain. To construct the CNN inputs, we apply both spatial and temporal downsampling: the fields are spatially downsampled to $64\times128$, and each case is temporally downsampled by a factor of $5$, reducing the original $1190$ snapshots to $238$ frames.

Due to the pronounced discrepancy in sample size between the two datasets, we adopt different data-splitting protocols to construct the training, validation, and test sets. For the cylinder-flow dataset, where the cases are sufficiently sampled, we perform a case-level split: we randomly select $10\%$ of the complete cases for validation and another $10\%$ for testing, with the remaining cases used for training. This means that the data used for testing belongs to entirely new simulations whose parameters were never seen by the network in the training phase. In contrast, the airfoil dataset is a small-sample dataset with sparse coverage of the physical-parameter space; a case-level split would make the evaluation highly sensitive to the particular partition. To avoid introducing additional confounding factors and to prevent information leakage, we instead apply an intra-case temporal split for the airfoil dataset: for each case, we extract two disjoint contiguous segments, each covering $10\%$ of the time series, as the validation and test sets, respectively, and use the remaining time steps for training.

\section{Results}
\label{sec:results}

This section presents an assessment of the proposed method from multiple perspectives, including its ability to capture the intrinsic low-dimensional manifold structure, to extract latent factors with clear physical meaning, and to reconstruct flow fields with high fidelity.
For comparison, we include: a linear autoencoder (\ac{PCA}),  a classical manifold learning algorithm (\ac{ISOMAP}), and a state-of-the-art  \ac{VAE} ($\beta$-\ac{VAE}), thereby highlighting the advantages brought by performing \ac{ELBO} decomposition within the \ac{VAE} framework. Furthermore, for the airfoil gust-encounter dataset we also qualitatively compare the obtained manifold structure with that obtained with an observation-augmented autoencoder~\cite{fukami2023grasping}. 
By including classical and state-of-the-art approaches in this field as baselines, we aim to provide a fair comparison and a comprehensive demonstration of the capabilities of the proposed method.

\subsection{Wake of the cylinder-in-channel}

The cylinder dataset is characterized by large-scale vortex shedding within the channel confinement, yielding flow fields that remain coherent and well-organized rather than chaotic. Our goal on this dataset is therefore to uncover a latent manifold with clear physical meaning.
Accordingly, within the DKL-VAE framework we keep the coefficients of the MI and Dim-KL terms at their original VAE ELBO values (i.e., 1), and only increase the weight of the TC term to encourage a more structured and physically interpretable latent space. We sweep the TC coefficient in the range $[1, 50]$ and observe that the latent manifold distribution becomes stable with $\lambda_{\mathrm{TC}} \gtrsim 10$. A large $\lambda_{\mathrm{TC}}$, on the other hand, may lead to degradation in reconstruction accuracy and can also introduce training instability. Based on this trade-off, we set $\lambda_{\mathrm{TC}} = 15$ for the results reported below. In addition, following the latent-dimension selection procedure described in $\S$~\ref{sec:elbo-decomp}, we determine that a $5$-dimensional latent space is sufficient for the cylinder dataset. Therefore, for a fair comparison, the latent dimensionality of all methods on the cylinder dataset is fixed to $5$.

Figure~\ref{fig:cyl-loss-curves} shows the training loss histories of $\beta$-\ac{VAE} and DKL-\ac{VAE}. Both methods are trained with the same network architecture and training protocol. In the case of the $\beta$-VAE, owing to its sensitivity to the choice of $\beta$ in terms of training stability and the resulting latent distribution, we set $\beta = 15$ to ensure consistency with the weight applied to the $\mathcal{L}_{\mathrm{TC}}$ term in the DKL-VAE. The same MSS estimator is employed to report the decomposed loss terms, while its training objective still uses the original formulation with the standard (non-decomposed) \ac{KL} regularization.

Figure \ref{fig:cyl-loss-curves}.a reports the behavior of the reconstruction loss $\mathcal{L}_{\mathrm{rec}}$ during the training and the validation for both $\beta$-VAE (in blue) and DKL-\ac{VAE} (in red), across the epochs. DKL-\ac{VAE} consistently exhibits lower values in both phases, as proof of the higher accuracy achieved. On the other hand,  MI loss $\mathcal{L}_{\mathrm{MI}}$ (Fig. \ref{fig:cyl-loss-curves}.b), TC loss  $\mathcal{L}_{\mathrm{TC}}$ (Fig. \ref{fig:cyl-loss-curves}.c) and Dim-KL loss $\mathcal{L}_{\mathrm{Dim\text{-}KL}}$ (Fig. \ref{fig:cyl-loss-curves}.d)  exhibit, at a first glance, similar performance in terms of disentanglement of the latent space. However, a slightly lower $\mathcal{L}_{\mathrm{MI}}$, a higher reconstruction loss $\mathcal{L}_{\mathrm{rec}}$, and a smaller $\mathcal{L}_{\mathrm{Dim\text{-}KL}}$ highlight a stronger regularization imposed by the $\beta$-VAE. These trends suggest that the latent representation of $\beta$-VAE retains less information about the data, while the marginal distribution is pushed more aggressively toward the Gaussian prior; such over-regularization can distort the intrinsic manifold structure. Notably, although ELBO-TC decomposition enables independent weighting of the individual terms, their optimization effects are usually not independent: the gradients induced by different components can be coupled, and the resulting loss trajectories may depend on dataset-specific characteristics.

Figure~\ref{fig:cyl-latent-space} shows the latent-space distributions obtained across the different methods. The latent dimensions are indexed according to their importance within each method: for PCA and ISOMAP, the dimensions follow the explained-variance ratio, while for $\beta$-VAE and DKL-VAE they are sorted by the latent activity defined in Eq.~\eqref{eq:kl-1d-closed}. 

Overall, all methods recover structured low-dimensional distributions. In particular, they all identify a limit-cycle structure associated with periodic vortex shedding, together with latent dimensions that correlate with the cylinder position parameters. However, the detailed manifold geometry and its correlation with physical parameters differ substantially across methods. Here, we construct the neighborhood graph using the $100$ nearest neighbors to perform ISOMAP, corresponding to the number of snapshots of 2 cases out of 1036.

PCA and ISOMAP produce broadly similar low-dimensional distribution shapes: the key difference is that the PCA embedding is noticeably tilted with respect to the latent coordinate axes, while ISOMAP's coordinates are better aligned with the embedding. $\beta$-VAE and DKL-VAE, in contrast, produce manifold geometries that are fundamentally different from those obtained by the classical methods. As shown in Fig.~\ref{fig:cyl-latent-space}(a1--d1), instead of the ring-like/cylindrical structure observed with PCA and ISOMAP, the learned latent distributions of $\beta$-VAE and DKL-VAE exhibit a spindle-shaped geometry: the radius is larger in the middle and gradually narrows toward both ends. This behavior is mainly induced by the disentanglement effect, under which the physical position of the cylnder within the channel $y_c$ becomes more separable along the latent dimension $z_1$. Consequently, samples corresponding to relatively larger or smaller $y_c$ (i.e., cylinders located closer to the upper or lower wall) are pulled toward the two ends of the manifold. The observed contraction of the limit-cycle trajectory near the tips reflects the influence of wall confinement.

The same disentanglement effect is also observed in Fig.~\ref{fig:cyl-latent-space} (a2--d2): both $\beta$-VAE and DKL-VAE learn latent distributions in which the cylinder position parameter $x_c$ becomes more separable along the $z_2$ dimension. However, 
the overly strong constraint induced by $\beta$ on the marginal distribution (i.e., an excessively large weight on $\mathcal{L}_{\mathrm{Dim\text{-}KL}}$) tends to distort and tilt the manifold geometry of the $\beta$-VAE. In contrast, thanks to the decomposition introduced in this paper, DKL-VAE is the only method that correctly disentangles and captures the trend whereby increasing $x_c$ leads to a slight contraction of the manifold radius, consistent with the observed reduction in diffusion effects.

The first two rows of Figure~\ref{fig:cyl-latent-alignment} further presents the projections of the learned latent representation onto the  spatial coordinates, showing how the latent variables vary with the cylinder size and position parameters. The normalized HSIC values between latent dimensions and physical parameters are annotated in the figure as quantitative indicators of their statistical dependence. As shown in the first row of the figure, the displayed latent dimension of $\beta$-VAE and DKL-VAE exhibits both a stronger correlation with $y_c$ and improved separability. These improvements are reflected in two observable patterns: (i) the distributions become increasingly compact, and (ii) samples from the same case collapse to an approximately single point rather than forming an elongated trace along that latent coordinate. Interestingly enough, DKL-VAE and PCA learn latent coordinates that correlate well with $x_c$, whereas DKL-VAE exhibits noticeably better separability. The most interesting finding from these manifold projections is that the DKL-VAE manifold not only achieves the strongest correlation with both $y_c$ and $x_c$, but also reveals a clear physical relationship as shown in Fig.~\ref{fig:cyl-latent-alignment}.d1: $y_c$ is approximately linearly related to the latent coordinate $z_1$, and the magnitude of the slope scales proportionally with the cylinder radius $R$. In other words, this single latent dimension simultaneously encodes the wall-confinement and wall-proximity effects jointly determined by the cylinder center location and its radius, while encompassing these nonlinear flow interactions into an explicit linear dependence in the latent space. The better consistency of the DKL-VAE latent space with the underlying physical parameters is also supported by its higher normalized HSIC values. It should be noted, however, that the nHSIC values in the first row quantify only the nonlinear dependence between the latent coordinates and $y_c$. They do not account for slope changes with the cylinder radius $R$.
Fig.~\ref{fig:cyl-latent-alignment}(a.3--d.3) further shows the pairwise nHSIC values among the latent variables. According to the results, the latent variables obtained by PCA and ISOMAP exhibit strong mutual coupling, indicating that different underlying factors and physical meanings remain heavily entangled. On the other hand, a comparison between $\beta$-VAE and DKL-VAE reveals a clearer difference in latent-space organization. For $\beta$-VAE, noticeable entanglement persists across multiple latent dimensions, particularly among $z_2$--$z_5$. For DKL-VAE, although the peak nHSIC values are stronger, they are concentrated only in a few latent-dimension pairs, mainly between $z_3$ and $z_4$. These two dimensions correspond to the limit-cycle structure, and their strong dependence is physically expected, as it reflects the intrinsic phase-angle correlation of the limit cycle. This suggests that DKL-VAE correctly captures the physically meaningful statistical dependence while achieving better disentanglement in the remaining latent dimensions.

\begin{table}[t]
\centering
\caption{Relative $\ell_2$ reconstruction error on the two datasets across different methods. For $\beta$-VAE and DKL-VAE, the reported results are obtained from five runs with different random seeds for data-batch partitioning and model initialization, together with the corresponding standard deviations.}
\label{tab:recon_error_two_datasets}
\begin{tabular}{lcc}
\hline
\textbf{Method} & \textbf{Cylinder dataset} & \textbf{Airfoil dataset} \\
\hline
PCA         &  $22.2\%$        & $59.5\%$      \\
Isomap      &  $19.4\%$        & $45.3\%$   \\
$\beta$-VAE &  $11.9\% \pm 0.09\%$        & $23.6\% \pm 0.26\%$  \\
DKL-VAE     &  $11.2\%  \pm 0.12\%$        & $23.2\% \pm 0.31\%$   \\
\hline
\end{tabular}
\end{table}

The relative $\ell_2$ error on the reconstruction of the $2D$ velocity fields over the entire domain is employed here to quantify the reconstruction accuracy, whose results are reported in Tab.~\ref{tab:recon_error_two_datasets}. Since ISOMAP does not provide an explicit inverse mapping, we reconstruct the high-dimensional field by taking the $5$ nearest neighbors in the training set and performing a distance-weighted interpolation in the original space. Overall, DKL-VAE achieves the lowest reconstruction error, being approximately half that of PCA and about $5\%$ lower relatively than $\beta$-VAE on the cylinder dataset.

Figure~\ref{fig:cyl-recon} highlights representative flow-field snapshots in which the reconstruction discrepancies among methods are more pronounced, together with the corresponding error contours obtained as the difference between the reconstructed and reference fields. For the first case shown in rows 1--4 of Fig.~\ref{fig:cyl-recon}, which lies in a typical parameter regime of unsteady vortex shedding, both PCA and ISOMAP exhibit noticeable deviations from the ground-truth flow field. Although they capture the presence of the shed vortex street, they fail to accurately reconstruct the local flow structures. In contrast, $\beta$-VAE and DKL-VAE recover the original flow features more faithfully and provide a reconstruction closer to the ground truth. For the second case shown in Fig. \ref{fig:cyl-recon} rows 5--8, the inlet velocity is relatively low and the flow is only slightly above the critical Reynolds number. The overall magnitude of the $v$-component is small but still shows weak unsteady behavior, which makes reconstruction more challenging. Among the four methods, ISOMAP and $\beta$-VAE collapse the reconstruction to a steady flow field, while PCA produces a spurious region with negative $v$ that is absent in the ground truth. Only DKL-VAE successfully reconstructs the underlying flow structures consistently with respect to the original field.

\subsection{Gust encountering airfoil NACA 0012}

The gust encountering airfoil dataset involves incoming large-scale vortical gusts that substantially reshape the vorticity field around the airfoil, inducing intense flow separation or stall, secondary vortices, and strong nonlinear interactions across multiple vortex scales. Due to the inherent irregularity and complexity of this testcase, we decided to increase the weight of $\mathcal{L}_{\mathrm{TC}}$ and additionally raise the weight of $\mathcal{L}_{\mathrm{Dim\text{-}KL}}$, to promote a more compact latent-space distribution while preserving the fidelity of the captured flow-field information. We conducted a parameter sweep over $\lambda_{\mathrm{TC}} \in [1, 50]$ and $\lambda_{\mathrm{Dim\text{-}KL}} \in [1, 20]$, and selected $\lambda_{\mathrm{TC}} = 10$ and $\lambda_{\mathrm{Dim\text{-}KL}} = 4$. In practice, the proposed method exhibits strong robustness to these hyperparameters, further discussed in this section. For this dataset, following the latent-dimension selection procedure described in $\S$~\ref{sec:elbo-decomp}, the required latent dimensionality is $9$. Therefore, in all subsequent analyses, we set the latent dimension to $9$ for all methods. 

Figure~\ref{fig:Airfoil-loss-curves} shows the convergence curves of the individual loss terms for $\beta$-VAE and DKL-VAE over training epochs. The two models exhibit very similar behaviors in $\mathcal{L}_{\mathrm{REC}}$ and $\mathcal{L}_{\mathrm{MI}}$, whereas the main discrepancy arises in $\mathcal{L}_{\mathrm{TC}}$ and $\mathcal{L}_{\mathrm{Dim\text{-}KL}}$. This can be attributed to the higher flow complexity, which makes reconstruction more challenging and leads to a significantly larger reconstruction loss than that of the cylinder dataset. Since $\mathcal{L}_{\mathrm{REC}}$ provides the dominant gradient during optimization, the training process tends to prioritize reconstruction, resulting in comparatively insufficient optimization of $\mathcal{L}_{\mathrm{TC}}$ and $\mathcal{L}_{\mathrm{Dim\text{-}KL}}$. By explicitly emphasizing $\mathcal{L}_{\mathrm{TC}}$, DKL-VAE achieves a lower degree of latent statistical dependence while avoiding overly strong constraints on $\mathcal{L}_{\mathrm{Dim\text{-}KL}}$.

The impact of different optimization trajectories is evident in the dominant three-dimensional latent-space structures, shown in Fig.~\ref{fig:Airfoiol-latent-space}. Here, \ac{ISOMAP} is performed using 200 nearest neighbors. As a baseline, the conventional methods fail to reveal a coherent manifold on this complex dataset. In the \ac{PCA} embedding, samples from different angles of attack are heavily entangled, while \ac{ISOMAP} separates the angles of attack into clusters enforcing consistency of the geodesic-distance metric in the low-dimensional embedding. Nonetheless, the resulting manifold remains highly irregular and disorganized. Compared with \ac{PCA} and \ac{ISOMAP}, the $\beta$-\ac{VAE} yields a more compact and structured latent manifold; however, noticeable distortion and skewness persist. In contrast, DKL-VAE reveals a clean and more interpretable geometry of the latent space. For the flow without gusts, the vortex shedding is represented as a limit cycle, and the limit cycles corresponding to different angles of attack are approximately parallel to each other. For gust-perturbed flows, variations in shedding strength and amplitude are encoded by in-plane expansions or contractions of the angle-of-attack limit-cycle plane, while the gust-induced instantaneous change in the effective angle of attack is captured by displacements of the trajectory along the normal direction to the limit-cycle plane.
\begin{figure}[t]
  \centering
  \begin{overpic}[width=0.7\columnwidth, unit = 1mm]{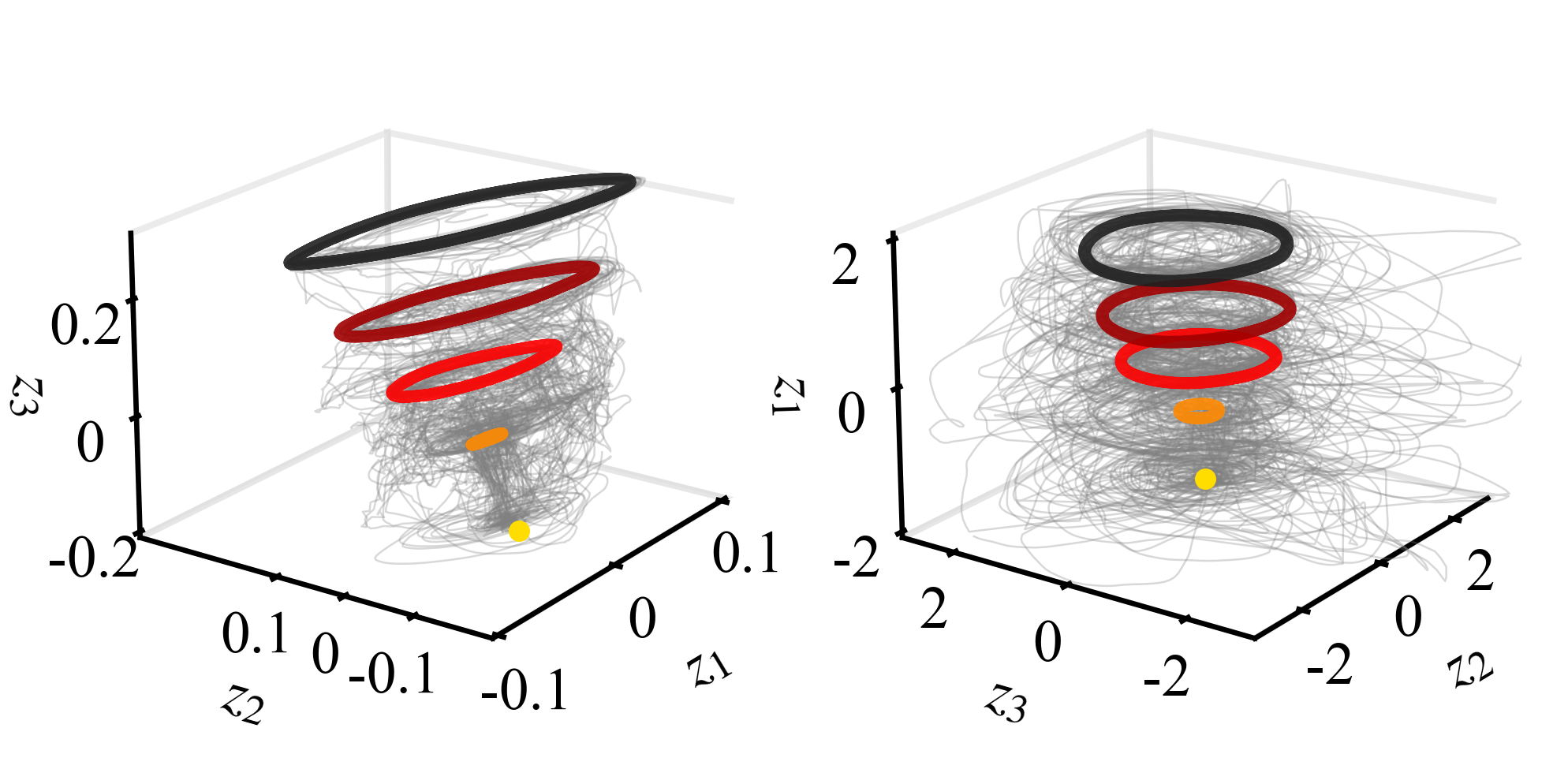}
\put(0,41){\parbox{10mm}{{a)}}}    
\put(50,41){\parbox{10mm}{{b)}}}   
    \put(91,39){\parbox{20mm}{\centering {$\alpha[\text{deg}]$}}}

\put(97,33){\roundedline{AOA20}\hspace{1mm}20}
\put(97,29){\roundedline{AOA30}\hspace{1mm}30}
\put(97,25){\roundedline{AOA40}\hspace{1mm}40}
\put(97,21){\roundedline{AOA50}\hspace{1mm}50}
\put(97,17){\roundedline{AOA60}\hspace{1mm}60}
  \end{overpic}
  \caption{Latent-space distributions on the airfoil dataset: (a) observation-augmented autoencoder, with the latent-space visualization replotted from the latent-coordinate  data of \citet{fukami2023grasping}; (b) DKL-VAE.}
  \label{fig:Airfoilcomp-Fukami}
\end{figure}

\begin{figure}[t]
  \centering
  \begin{overpic}[width=0.7\columnwidth, unit = 1mm]{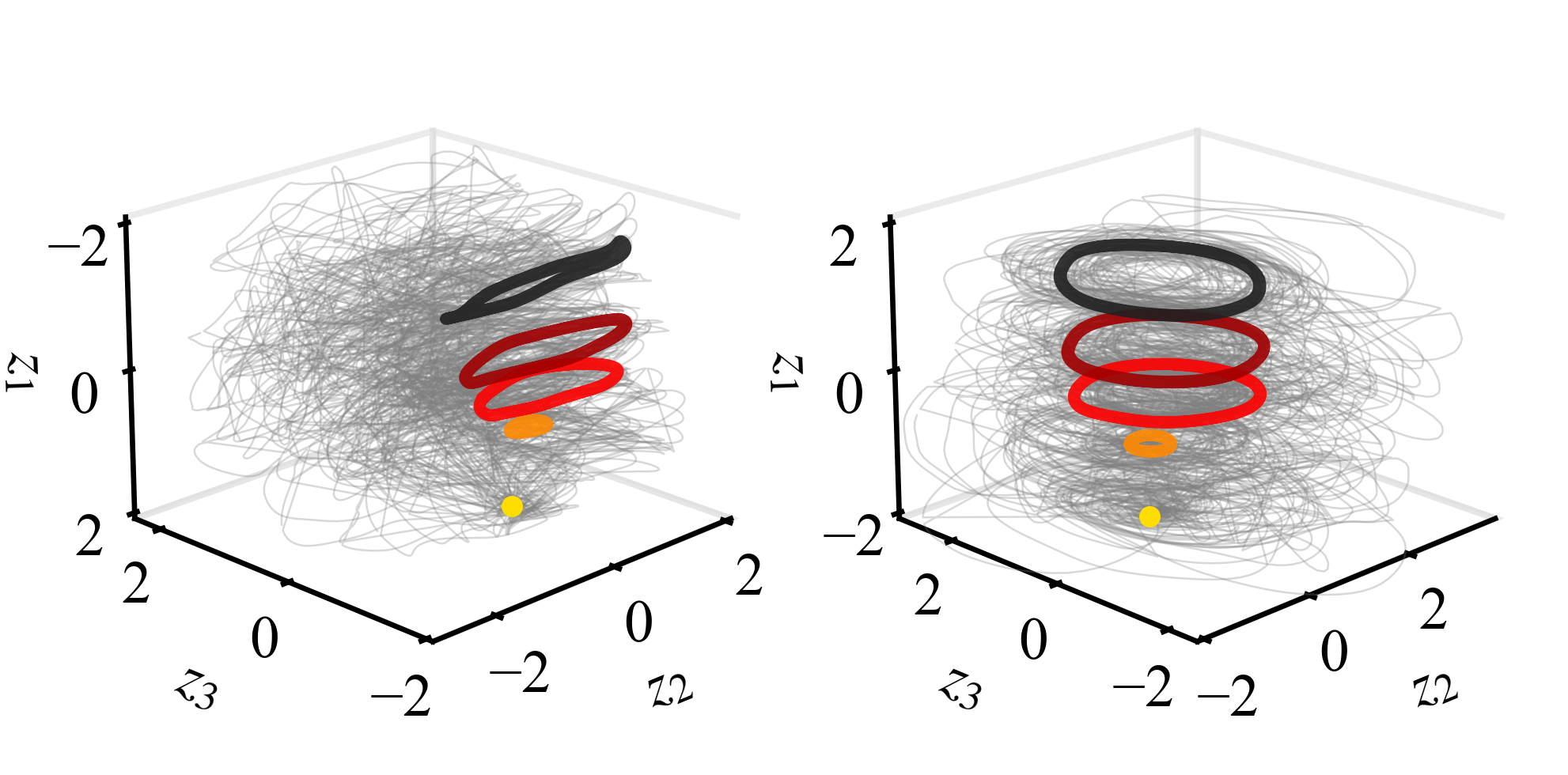}
\put(0,41){\parbox{10mm}{{a)}}}    
\put(50,41){\parbox{10mm}{{b)}}}   
    \put(91,39){\parbox{20mm}{\centering {$\alpha[\text{deg}]$}}}

\put(97,33){\roundedline{AOA20}\hspace{1mm}20}
\put(97,29){\roundedline{AOA30}\hspace{1mm}30}
\put(97,25){\roundedline{AOA40}\hspace{1mm}40}
\put(97,21){\roundedline{AOA50}\hspace{1mm}50}
\put(97,17){\roundedline{AOA60}\hspace{1mm}60}
  \end{overpic}
  \caption{Loss-weight robustness verification: (a) latent space of $\beta$-VAE with $\beta=40$; (b) latent space of DKL-VAE with $\lambda_{\mathrm{TC}}=40$ and $\lambda_{\mathrm{Dim\text{-}KL}}=16$.}
  \label{fig:Airfoilparametric-val}
\end{figure}
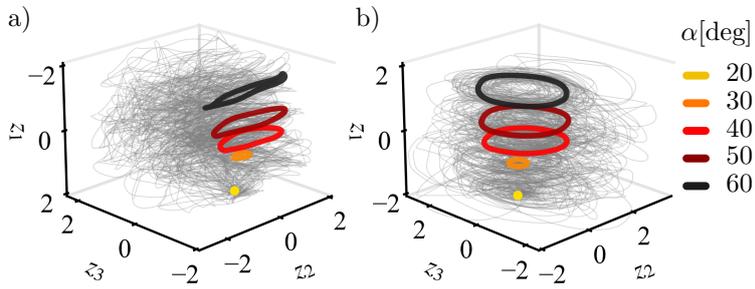

Fig.~\ref{fig:Airfoil-latent-alignment}(a.1--d.1) further compares the correlation between the learned latent representation and the lift coefficient $C_l$. In the \ac{PCA}, \ac{ISOMAP}, and $\beta$-\ac{VAE} embeddings, the lift coefficient values exhibit some degree of clustering on top of the scattered latent distributions; however, such patterns are difficult to interpret in terms of physically meaningful variables. In contrast, in the latent representation learned by DKL-VAE, the lift shows a clear and regular correlation with the $z_1$ coordinate, which provides an informative indicator of the effective angle of attack: each nominal angle of attack corresponds to its own approximately parallel orbital plane. During phases where the airfoil does not strongly interact with the gust, the trajectory remains inside the corresponding orbital plane, associated with moderate lift levels. When the airfoil encounters a gust with aerodynamic characteristics markedly altered, the trajectory departs from the orbital plane and shifts along the normal direction; the magnitude of this normal displacement is positively correlated with the instantaneous lift variation. From the normalized HSIC results in Fig.~\ref{fig:Airfoil-latent-alignment}(a.2--d.2), a trend similar to that observed for the cylinder dataset can be identified. The latent dimensions obtained by PCA and ISOMAP remain strongly entangled, indicating that different underlying factors are not clearly separated. In comparison, $\beta$-VAE shows relatively weak statistical dependences distributed across many latent dimensions, whereas for DKL-VAE the nHSIC values are mainly concentrated in only a few pairs of higher-order latent dimensions. Meanwhile, the lower-order dimensions, especially $z_1$--$z_{13}$, are significantly more disentangled from the other latent variables. The flow is characterized by complex coupling among the different stages of the gust--airfoil response, as well as the associated vortex-shedding dynamics. As a result, the higher-order latent dimensions in the airfoil dataset display more complex dependence relationships than those in the cylinder dataset.

It is worth noting that \citet{fukami2023grasping}, by introducing the instantaneous lift as an additional regularization for the same physical problem, extracted a similar dominant latent-space structure, as replotted from their latent-coordinate data in Fig.~\ref{fig:Airfoilcomp-Fukami}.a. The lift-based regularization tends to contract the embedding of the disturbed cases by bringing together samples with similar instantaneous lift, even when their underlying flow fields differ significantly. Another key difference, however, is that in the latent coordinates produced by their observation-augmented autoencoder, the coordinates are not aligned with the planes of the limit cycle and the axis of the cone, indicating that the effective angle-of-attack dimension might be coupled with the two limit-cycle dimensions. In our results, by contrast, the limit cycle is represented exclusively by the $z_2$--$z_3$ subspace, while the effective angle of attack is encoded solely along the $z_1$ direction, which is orthogonal to the $z_2$--$z_3$ plane. This clean separation assigns different physical effects to different latent coordinates, confirming higher disentanglement and interpretability. Moreover, this result is achieved purely leveraging the intrinsic properties of the latent space, without introducing any additional information as supervision; the dimensionality reduction and manifold learning remain fully unsupervised and data-driven.

As in the previous test case, the reconstruction accuracy is evaluated using the vorticity-based relative error, whose results are reported in Tab.~\ref{tab:recon_error_two_datasets}. The relative ranking among methods is largely consistent with that observed on the cylinder dataset; however, owing to the more intricate flow dynamics and the broader parameter variations in the gust encountering airfoil dataset, the overall reconstruction errors are generally higher. Nevertheless, DKL-VAE still achieves the highest reconstruction accuracy, yielding errors that are approximately half those of \ac{PCA} and slightly lower than those of $\beta$-\ac{VAE}. Figure~\ref{fig:Airfoil-recon} highlights representative cases from different parameter regimes where the reconstruction differences among methods are particularly pronounced. The differences among methods become, as expected, most pronounced under strong-gust conditions, where the flow field is highly complex. By comparing the reconstructed fields, it can be observed that \ac{PCA} is easily affected by the near-wall strong structures across different angles of attack, leading to visible non-smooth footprints from other angles of attack. This low order reconstruction consistently filters out the biggest gradients induced by the gusts. The \ac{ISOMAP} reconstructions, while recovering a bit more of these peaks, still heavily depend on the availability of close neighbors and are therefore also susceptible to contamination from other gust conditions and angles of attack. For instance, the fields in Fig.~\ref{fig:Airfoil-recon}.b3 and Fig.~\ref{fig:Airfoil-recon}.b5 exhibit clear interpolation toward the near-wall vorticity patterns associated with different angles of attack, while Fig.~\ref{fig:Airfoil-recon}.b1 and Fig.~\ref{fig:Airfoil-recon}.b7 are interpolated toward flows under different gust settings, where both the gust location and its intensity deviate noticeably from the target field. These discrepancies are more clearly revealed in the corresponding error contours. In comparison, $\beta$-VAE and DKL-VAE yield overall better reconstructions. Although mild edge non-smoothness appears in regions with abrupt vorticity variations, they successfully recover the dominant near-wall and wake structures of the flow. It can also be observed that the reconstruction accuracy of DKL-VAE does not show a significant improvement over that of $\beta$-VAE. It must be remarked that a conventional autoencoder trained exclusively with a reconstruction loss would generally achieve better performance. However, the DKL-VAE aims at preventing the latent-space structure from being excessively distorted by overly strong dimension-wise KL regularization, as may occur in $\beta$-VAE, while retaining a good reconstruction accuracy. By leveraging the ELBO decomposition, the proposed method promotes a more physically consistent and interpretable latent space while maintaining competitive reconstruction performance.

\subsection{Sensitivity to weight tuning}

Since our approach decomposes the single KL term in the VAE ELBO and assigns separate weights to the resulting components, a natural concern is that the additional hyperparameters may increase tuning effort and introduce sensitivity to weight choices. In practice, however, we find that the proposed method exhibits strong weight robustness on both datasets considered in this work: across a wide range of weights, it consistently recovers the correct latent manifold structure. An additional observation is that, when weighting multiple decomposed terms, the \emph{ratios} between weights, rather than their absolute magnitudes, play a more decisive role in shaping the latent distribution.This implies that the effective degree of freedom in weight tuning, to a large extent, remains one, and thus does not introduce more free hyperparameters than $\beta$-VAE. To illustrate these two points, we present results on the gust encountering airfoil dataset with all weights increased by a factor of four. Specifically, we set $\beta=40$ for $\beta$-VAE, and $\lambda_{\mathrm{TC}}=40$ and $\lambda_{\mathrm{Dim\text{-}KL}}=16$ for DKL-VAE. As shown by the embeddings in Fig.~\ref{fig:Airfoilparametric-val}, the latent space of $\beta$-VAE (Fig.~\ref{fig:Airfoilparametric-val}.a) becomes more distorted under the stronger penalty and tends to collapse toward the Gaussian prior, yielding an embedding that is closer to a spherical cloud compared with Fig.~\ref{fig:Airfoiol-latent-space}. In contrast, DKL-VAE (Fig.~\ref{fig:Airfoilparametric-val}.b) largely preserves the learned manifold geometry: it still clearly captures the limit-cycle orbits as well as the disentangled effective angle-of-attack coordinate. The main difference from Fig.~\ref{fig:Airfoiol-latent-space} is a mild local stretching/compression that makes the distribution appear slightly more compact and regular.

Based on our experience with the two datasets considered in this work, we suggest a sequential hyperparameter-tuning protocol: first tune $\lambda_{\mathrm{TC}}$, then tune $\lambda_{\mathrm{Dim\text{-}KL}}$. A practical heuristic is to inspect embeddings produced by conventional methods such as \ac{PCA} or \ac{ISOMAP}: if the latent space appears irregular or severely distorted, as in Fig.~\ref{fig:Airfoiol-latent-space}(a,b), one may increase $\lambda_{\mathrm{Dim\text{-}KL}}$ to promote a more geometrically regular representation. 

While $\lambda_{\mathrm{MI}}$ is fixed to $1$ as in the standard \ac{VAE} for the datasets tested in this paper, it may be increased for highly noisy data or when more aggressive filtering is desired. Note that, although the decomposed terms are formally separated, their optimization effects can still be coupled. In principle, these weights can be set automatically via adaptive loss-balancing methods~\citep{Vandenhende2022MTLSurvey} to learn problem-specific trade-offs among objective terms; such strategies have recently been incorporated into \ac{VAE}-based models (e.g., learnable $\beta$) with promising results~\citep{Ozcan2025LVAE}.

\section{Discussion and Conclusions}
\label{sec:conclusions}

In this work, we introduce an information-theoretic variational autoencoder framework to manifold learning for fluid flows. The main idea is to adopt the ELBO-TC decomposition, which splits the KL-divergence term in the standard VAE ELBO into physically interpretable components: mutual information, total correlation, and dimension-wise KL terms.
This decomposition enables targeted design and regularization of latent-space properties, facilitating controllable and physically meaningful representations. Compared with approaches that impose a single strong KL penalty on the original ELBO (e.g., $\beta$-VAE), it mitigates the undesirable effects of aggressively pulling complex data distributions toward a noninformative simple prior.

The proposed method is validated on two challenging test cases, such as the flow in a channel with a cylinder located at various positions and a gust-encountering airfoil at large angles of attack, and benchmarked against classical manifold-learning approaches (PCA and ISOMAP) and state-of-the-art deep-learning baselines ($\beta$-VAE and observation-augmented autoencoders). 
The results demonstrate that the proposed method effectively captures latent representation with clear physical interpretability, yielding low-dimensional coordinates that disentangle distinct physical effects while achieving the best flow-field reconstruction accuracy among the compared methods. Although not explicitly defined in a strict sense, the method captures and well approximates the physical manifolds induced by the intrinsic flow dynamics, thereby serving as a manifold-learning approach. In addition, the proposed framework exhibits good robustness to hyperparameter choices, reducing the practical burden of weight tuning. 

Along with its role as a latent-space regularization for reduced-order representation, the explicit mutual information and total correlation decomposition in the probabilistic framework also provides a principled way to separate and distill information in the flow state, thereby establishing a direct link to information-theoretic analyses of turbulent flows. Recently, \citet{arranz2024informative} introduced an \emph{informative and non-informative decomposition} of turbulent channel-flow velocity fields with respect to the wall-shear stress. Within the present information-decomposition variational-autoencoder framework, similar objectives can be achieved by jointly optimizing constraints on the total correlation among latent variables and the mutual information between selected latent factors and the physical quantity of interest. This suggests that the proposed framework can be extended to a broader range of information-theoretic analyses of complex flows.  Beyond fundamental studies of fluid physics, autoencoder-based dimensionality reduction methods also hold significant potential for aircraft design applications. For example, low-dimensional representations can support surrogate aerodynamic prediction~\cite{FrancesBelda2024BetaVAE,zhang2024heterogeneous} and compact shape parameterization for generative inverse design~\cite{Wang2025GFE}. The proposed DKL-VAE method demonstrates improved physical interpretability through the correspondence between latent variables and relevant physical parameters, making it also a promising tool for aerodynamic surrogate modeling and inverse design.

Overall, the proposed approach provides a powerful tool for manifold learning in high-dimensional, nonlinear flow systems, easy to implement in common variational frameworks. Future work will include validating the proposed method on more challenging high-Reynolds-number flows with measurement noise, exploiting informative statistics, and exploring its use in low-dimensional-representation-based physics discovery, efficient flow sensing and control.

\section*{Appendix: ELBO-TC decomposition derivation}

This appendix derives the decomposition of the index-averaged KL term used in Eq.~\eqref{eq:elbo-tc}. Following Appendix C of Ref.~\cite{chen2018isolating}, we show how the KL regularizer can be written as the sum of three terms with distinct information-theoretic meanings: index-code mutual information, total correlation, and dimension-wise KL divergence.


\vspace{0.4cm}

\textbf{Step 1. Rewriting the dataset average as an expectation.}

Let $n \sim p(n)$ denote a uniformly sampled data index:
\begin{equation}
    p(n)=\frac{1}{N}, \qquad n\in\{1,\dots,N\}.
\end{equation}
For consistency with the main text, we denote
\begin{equation}
    q_\phi(\bm{z}\mid n)\equiv q_\phi(\bm{z}\mid \bm{x}_n).
\end{equation}
Then the index-averaged \ac{KL} term can be written as:
\begin{equation}
    \frac{1}{N}\sum_{n=1}^N \mathrm{KL}\big(q_\phi(\bm{z}\mid n)\,\|\,p(\bm{z})\big)
    =
    \mathbb{E}_{p(n)}\!\left[
    \mathrm{KL}\big(q_\phi(\bm{z}\mid n)\,\|\,p(\bm{z})\big)
    \right].
    \label{eq.app1}
\end{equation}
The aggregated posterior can be defined as
\begin{equation}
    q_\phi(\bm{z})=\sum_{n=1}^N p(n)\,q_\phi(\bm{z}\mid n),
\end{equation}
which is the marginal of the joint distribution:
\begin{equation}
    q_\phi(\bm{z},n)=q_\phi(\bm{z}\mid n)\,p(n).
\end{equation}

\vspace{0.2cm}

\textbf{Step 2. Expanding the KL divergence.}

By definition,
\begin{equation}
    \mathrm{KL}\big(q_\phi(\bm{z}\mid n)\,\|\,p(\bm{z})\big)
    =
    \mathbb{E}_{q_\phi(\bm{z}\mid n)}
    \left[
    \log q_\phi(\bm{z}\mid n)-\log p(\bm{z})
    \right].
\end{equation}
Substituting this into Eq.~\eqref{eq.app1} gives
\begin{equation}
\begin{aligned}
\mathbb{E}_{p(n)}\bigl[
\mathrm{KL}\bigl(q_\phi(\bm{z}\mid n)\,\|\,& p(\bm{z})\bigr)
\bigr]
 = {}  \\
& \hfill \mathbb{E}_{q_\phi(\bm{z},n)}\bigl[
\log q_\phi(\bm{z}\mid n)-\log p(\bm{z})
\bigr].
\end{aligned}
\label{eq.app2}
\end{equation}

\vspace{0.2cm}

\textbf{Step 3. Adding and subtracting auxiliary terms.}

With the factorized prior $p(\bm{z})=\prod_j p(z_j)$, we add and subtract
$\log q_\phi(\bm{z})$ and $\log \prod_j q_\phi(z_j)$ inside the expectation:
\begin{equation}
\begin{aligned}
    \mathbb{E}_{q_\phi(\bm{z},n)}\!\Big[
    &\log q_\phi(\bm{z}\mid n)-\log q_\phi(\bm{z}) \\
    &+\log q_\phi(\bm{z})-\log \prod_j q_\phi(z_j) \\
    &+\log \prod_j q_\phi(z_j)-\log \prod_j p(z_j)
    \Big].
\end{aligned}
\label{eq.app3}
\end{equation}

\vspace{0.2cm}

\textbf{Step 4. Regrouping the terms.}

Regrouping Eq.~\eqref{eq.app3} yields:
\begin{equation}
\begin{aligned}
    &\mathbb{E}_{q_\phi(\bm{z},n)}\!\left[
    \log\frac{q_\phi(\bm{z}\mid n)}{q_\phi(\bm{z})}
    \right] \\
    &\quad+
    \mathbb{E}_{q_\phi(\bm{z})}\!\left[
    \log\frac{q_\phi(\bm{z})}{\prod_j q_\phi(z_j)}
    \right] \\
    &\quad+
    \mathbb{E}_{q_\phi(\bm{z})}\!\left[
    \sum_j \log\frac{q_\phi(z_j)}{p(z_j)}
    \right].
\end{aligned}
\label{eq.app4}
\end{equation}
In the second and third terms, the expectation is taken over $q_\phi(\bm{z})$ because the integrands do not depend on $n$.

\vspace{0.2cm}

\textbf{Step 5. Recognizing the three terms.}

Using $q_\phi(\bm{z},n)=q_\phi(\bm{z}\mid n)p(n)$, the first term becomes:
\begin{equation}
\begin{aligned}
    \mathbb{E}_{q_\phi(\bm{z},n)}\!\left[
    \log\frac{q_\phi(\bm{z}\mid n)}{q_\phi(\bm{z})}
    \right]
    &=
    \mathbb{E}_{q_\phi(\bm{z},n)}\!\left[
    \log\frac{q_\phi(\bm{z},n)}{q_\phi(\bm{z})p(n)}
    \right] \\
    &=
    \mathrm{KL}\big(q_\phi(\bm{z},n)\,\|\,q_\phi(\bm{z})p(n)\big).
\end{aligned}
\label{eq.app5}
\end{equation}
The third term can be reduced by marginalizing over $\bm{z}_{\setminus j}$:
\begin{equation}
\begin{aligned}
    \sum_j \mathbb{E}_{q_\phi(\bm{z})}\!\left[
    \log\frac{q_\phi(z_j)}{p(z_j)}
    \right]
    =
    \sum_j \mathrm{KL}\big(q_\phi(z_j)\,\|\,p(z_j)\big).
\end{aligned}
\label{eq.app6}
\end{equation}
Similarly, the second term is exactly the total correlation:
\begin{equation}
    \mathbb{E}_{q_\phi(\bm{z})}\!\left[
    \log\frac{q_\phi(\bm{z})}{\prod_j q_\phi(z_j)}
    \right]
    =
    \mathrm{KL}\!\left(
    q_\phi(\bm{z})\,\Big\|\,\prod_j q_\phi(z_j)
    \right).
\label{eq.app7}
\end{equation}

\vspace{0.2cm}

\textbf{Step 6. Final decomposition.}

Combining Eqs.~\eqref{eq.app4}--\eqref{eq.app7}, leads to:
\begin{equation}
\begin{aligned}
\frac{1}{N}\sum_{n=1}^N
\mathrm{KL}\big(q_\phi(\bm{z}\mid n)\,\|\,p(\bm{z})\big)
&=
\underbrace{\mathrm{KL}\big(q_\phi(\bm{z},n)\,\|\,q_\phi(\bm{z})p(n)\big)}_{\text{(i) Index-Code MI}} \\
&\quad+
\underbrace{\mathrm{KL}\!\left(
q_\phi(\bm{z})\,\Big\|\,\prod_j q_\phi(z_j)
\right)}_{\text{(ii) Total Correlation}} \\
&\quad+
\underbrace{\sum_j \mathrm{KL}\big(q_\phi(z_j)\,\|\,p(z_j)\big)}_{\text{(iii) Dimension-wise KL}}.
\end{aligned}
\label{eq.app8}
\end{equation}
This is exactly the decomposition used in Eq.~\eqref{eq:elbo-tc} in the main text with $q_\phi(\bm{z})$ denoting the aggregated posterior.

\section*{Acknowledgements}
This work was partially supported by the project EXCALIBUR (Grant No PID2022-138314NB-I00), funded by MCIU/AEI/10.13039/501100011033 and by“ERDF A way of making Europe”, as well as by the project SPANDRELS (SParse AND paRsimonious Event-based fLow Sensing), which has received funding from the European Union’s Horizon Europe research and innovation program under grant agreement No 101171280 (ERC-2024-COG). Views and opinions expressed are however those of the authors only and do not necessarily reflect those of the European Union or the European Research Council. Neither the European Union nor the granting authority can be held responsible for them.

\section*{Data availability statement}
The code used in this work is made available at: \url{https://github.com/Howlett-ZW/DKL-VAE}. The dataset is openly available in Zenodo, accessible through the
following link: \url{https://doi.org/10.5281/zenodo.19235606}.

\section*{Declaration of generative AI and AI-assisted technologies in the manuscript preparation process}

During the preparation of this work the author(s) used ChatGPT in order to improve the readability and language of specific sentences of the manuscript. After using this tool/service, the author(s) reviewed and edited the content as needed and take(s) full responsibility for the content of the published article.

\bibliography{sn-bibliography}

\end{document}